\documentclass[12pt]{article}
\usepackage{array}
\usepackage{tabularx}
\usepackage{amsmath,amssymb,amsfonts,mathrsfs,bbm,epsfig,cancel,cite,setspace,bigstrut,framed,eufrak,tensor,mathtools}
\usepackage{color}
\usepackage{pifont}
\usepackage{hyperref}
\usepackage{scalerel}
\usepackage[bbgreekl]{mathbbol}
\newcommand\fat[1]{\ThisStyle{\ooalign{%
  \kern.46pt$\SavedStyle#1$\cr\kern.33pt$\SavedStyle#1$\cr%
  \kern.2pt$\SavedStyle#1$\cr$\SavedStyle#1$}}}

\makeatletter \@addtoreset{equation}{section} \makeatother
\newcommand{\comment}[1]{}

\newcommand{\bw}{{\mathbf w}}

\newcommand{\bD}{{\mathbf D}}

\newcommand{\cM}{{\cal M}}

\newcommand{\cR}{{\cal R}}
\newcommand{\cS}{{\cal S}}

\newcommand{\cK}{{\cal K}}
\newcommand{\cL}{{\cal L}}

\newcommand{\cO}{{\cal O}}
\newcommand{\cQ}{{\cal Q}}

\newcommand{\cC}{{\cal C}}

\newcommand{\cF}{{\cal F}}
\newcommand{\cA}{{\cal A}}
\newcommand{\cU}{{\cal U}}
\newcommand{\cV}{{\cal V}}

\newcommand{\IR}{\mathbb{R}}

\newcommand{\ch}{\mathbb{ch}}

\newcommand{\iT}{\mathbbm{T}}

\newcommand{\ii}{\mathbbm{i}}
\newcommand{\iJ}{\mathbbm{J}}

\newcommand{\sD}{\mathscr{D}}

\newcommand{\bi}{\begin{itemize}}
\newcommand{\ei}{\end{itemize}}
\newcommand{\beq}{\begin{equation}}
\newcommand{\eeq}{\end{equation}}
\newcommand{\bea}{\begin{eqnarray}}
\newcommand{\eea}{\end{eqnarray}}

\DeclareFontEncoding{LS1}{}{}
\DeclareFontSubstitution{LS1}{stix}{m}{n}
\DeclareSymbolFont{stixletters}{LS1}{stix}{m}{it}
\DeclareMathAccent{\cev}{\mathord}{stixletters}{"91}
\DeclareMathAccent{\vec}{\mathord}{stixletters}{"92}
\DeclareMathAccent{\vecev}{\mathord}{stixletters}{"95}

\makeatletter
\newsavebox{\@brx}
\newcommand{\llangle}[1][]{\savebox{\@brx}{\(\m@th{#1\langle}\)}%
  \mathopen{\copy\@brx\kern-0.5\wd\@brx\usebox{\@brx}}}
\newcommand{\rrangle}[1][]{\savebox{\@brx}{\(\m@th{#1\rangle}\)}%
  \mathclose{\copy\@brx\kern-0.5\wd\@brx\usebox{\@brx}}}
\makeatother

\begin{document}

\begin{titlepage}
\vfill
\begin{flushright}
{\tt\normalsize KIAS-P24066}\\

\end{flushright}
\vfill
\begin{center}
{\Large\bf Lie, Noether, Kosmann, and\\ \vskip 3mm
Diffeomorphism Anomalies Redux}

\vskip 1.5cm

Taeyeon Kim$^\dagger$ and Piljin Yi$^\ddagger$
\vskip 5mm
$^\dagger${\it Department of Physics and Astronomy, Seoul National University, \\1 Gwanak-ro, Gwanak-gu, Seoul 08826, Korea}
\vskip3mm
$^\ddagger${\it School of Physics,
Korea Institute for Advanced Study, \\85 Hoegi-ro, Dongdaemun-gu, Seoul 02455, Korea}

\end{center}
\vfill

\begin{abstract}
The Noether procedure carries an inherent ambiguity due to the necessary local extension,
no longer a symmetry, of the global symmetry. The gauging should fix the ambiguity once
and for all, however, and, for translations, the general covariance demands us to use
the Lie derivative. We argue that, with this alone and without any further tweaking,
the Noether energy-momentum $\hat{\mathbbm{T}}$ must equal the symmetric
counterpart, $T$, inevitably and show the equality explicitly for general tensors.
For spinors, a subtlety with the Lie derivative itself enters the issue and leads us to the Kosmann lift,
often unnoticed by the physics community, from which $T=\hat{\mathbbm{T}}$ again emerges
straightforwardly and in a naturally symmetric form. Finally, we address how the same
Kosmann lift affects the anomaly computations and show that the diffeomorphism anomaly
from the seminal paper {\it must be halved} while the venerable anomaly polynomials
themselves stand unaffected. We discuss the ramifications of these findings.

\end{abstract}

\vfill
\end{titlepage}

\section{Introduction}

In this note, we come back to the age-old matter of the energy-momentum tensor, the
conservation law thereof, and the (anomalous) Ward identity for the diffeomorphism,
with fresh looks at what should be introductory quantum field theory
material.\footnote{References for these are too numerous and diverse to list,
in this note we will confine ourselves to the most immediate ones for our purpose.
The standard text we start with is the textbook by Weinberg \cite{Weinberg}.}
Our rationale for doing this is two-fold.

One is a long-standing confusion, resolved repeatedly via various different
approaches in the past, over the Noether \cite{Noether:1918} energy-momentum and on-and-off
appearance of the ``improvement"  terms \cite{Belinfante:1940,Rosenfeld:1940}.
We will offer a simple and clarifying view on the matter,
merely based on the general covariance, which  offers a sensible and
universal view on the nature of the Noether procedure itself. The key is how the
Lie derivative must be employed for the variation, and then how the Lie derivative
should act on spinors.

The other is the matter of the energy-momentum tensor of spinors and the Ward
identity thereof. As the role of Lie derivative is emphasized, an obvious
question to ask is how the diffeomorphism acts on spinors. We will invoke
the so-called Kosmann lift \cite{Kosmann:1971,Kosmann:1966-1,Kosmann:1966-2,Kosmann:1967}
and make an inevitable generalization for spinors coupled to gauge fields, and then revisit
the problem of general covariance and the energy-momentum tensors
for a very general theory of fermions. All these force us to revisit the diffeomorphism
anomaly and allow us to discover a subtle factor 1/2, to be delineated later.

Although the Noether procedure effectively produces the conserved current
related to a global and internal symmetry, the standard references leave much to
be desired when it comes to the energy-momentum tensor, with various different
approaches in existence. The simplest
way to bypass such is often via the gauging of the symmetry in question, whereby the
conserved current arises much more simply from the variation of the gauge field
or of the metric. It is in fact the latter version of the currents whose conservation
law elevates to the Ward identity most straightforwardly. This view is in part
responsible for why, even though the classical conservation laws require the equation
of motion, the Ward identity need not assume any such.

For the purpose of this note, we will distinguish the latter type of conserved current
with the notation $J$ and $T$, respectively to be called the gauge current and the symmetric
energy-momentum tensor, to be formally distinguished from their Noether counterpart
$\iJ$ and $\iT$. For fear of potential confusion, we will also introduce the notations
$\iJ\rightarrow \hat\iJ$ and $\iT\rightarrow \hat\iT$ as the respective covariantized versions.
At the end of the most naive version of the Noether procedure, an
``improvement" of $\hat\iT$ toward $T$ is then often invoked, although by tweaking
the Noether procedure, for instance in the case of  Yang-Mills field, to conform with
the internal gauge symmetries, $T=\hat\iT$ can be also achieved.

In the first one-third of this note, we will dissect this general set-up with the
gauging of the symmetry in question, dynamical or external, and offer a simple governing principle that would
give $\hat\iJ=J$ and $\hat\iT=T$, with no extra effort. For $T=\hat\iT$, in particular,
the central ingredient is how we must use the Lie derivative for the Noether variation
of matter fields,
\bea
\boldsymbol\delta_\xi \Phi=\mathfrak L_\xi \Phi\ ,
\eea
for any field $\Phi$. Although this looks like such a natural thing to do, to begin
with, we find few treatises of Noether energy-momentum under this rule,
strangely enough. We need to emphasize that, although we often encounter the
Lie derivative in the context of General Relativity, the operation has nothing
to do with the spacetime curvatures. For instance, the simple translational isometries of
the Minkowski spacetime must be written via a Lie derivative, for general tensors,
if we employ a coordinate system other than the Cartesian one.

With this choice, the crucial, if somewhat trivial, observation is that  the Lagrangian
$d$-form density $\cV\cL$, with the volume form $\cV$, transforms as
\bea
\boldsymbol\delta_\xi (\cV\cL)=d\left(\xi\lrcorner\,\cV\cL\right) \ .
\eea
The crux of the matter for us resides in the vanishing rest, which should be
subsequently split into two mutually canceling parts, one proportional to $T$
and the other to $\hat\iT$. From this observation, the equality of these two therefore
comes about almost trivially and without resorting to the equation of motion or
any other tweaking, as we will show repeatedly. Things get a little
more involved when the kinetic term of the matter field involves
the presence of the Levi-Civita connection explicitly, as with some higher-spin (bosonic)
fields, but we show that the general idea continues to hold.

All these thoughts then force us to ask the question of how the Lie derivative
should act on fermions. Spinors can be really defined properly only after we turn to
the orthonormal frames since the underlying Clifford algebra makes sense
with the local Lorentz indices. The Riemannian geometry may be reformulated with
the vielbein $e^a_{\;\,\mu}$ and the spin connection $w^{\;\;a}_{\mu \;\;b}$,
which brings us to the Kosmann lift of the Lie derivative and its action on
the local Lorentz frame and on spinors. In the middle one-third of this note,
we will spend much time on the related matter and then repeat the exercise that shows
$T=\hat\iT$ for spinors and offers the universal formulae for $T$ and $\iT$ separately.

The Kosmann lift refers to how the Lie derivative with its natural action on
tensors must be elevated to the frame bundle and to the spinor bundle.
A key fact relevant for the definition and computation of the symmetric energy-momentum
$T$ is that the Lie derivative of the vielbein under a vector field $\xi$  takes the following
form
\bea
\boldsymbol\delta_\xi e^a_{\;\,\mu} = \mathscr L_\xi e^a_{\;\,\mu} = {D^{(a} \xi^{b)} e_{b\mu}} \ .
\eea
On the other hand, for Noether energy-momentum $\iT$, the Kosmann lift on spinors works as
\bea
\mathscr{L}_\xi \Psi
=  \xi^\mu\partial_\mu \Psi -\frac14 \hat\xi^{ab}_{K}\gamma_{ab}\Psi    \ ,
\eea
for some antisymmetric rank-2 object $\hat\xi^{ab}_{K}$ computed from $\xi^\mu$
and the spin connection. The same operation may also be written in a more
manifestly covariant form,
\bea\label{KL_cov}
\mathscr{L}_\xi \Psi
=  \xi^\mu\mathscr D_\mu \Psi -\frac14 \hat\xi^{ab}_{V}\gamma_{ab}\Psi\ ,
\eea
with $\hat\xi^{ab}_{V}\equiv -D^{[a}\xi^{b]}$, from which  $\hat\xi^{ab}_{K}$ can be read
off.\footnote{Here we are assuming that $\Psi$ is not coupled to other gauge fields, but later
this shall be generalized to a fully covariant form with gauge fields.  }

If we wish to have a globally well-defined Lie derivative, over the frame bundle and the
spinor bundle, this Kosmann lift is unavoidable although, for some reason,
this fact is not widely appreciated by the physics community. The alternative, which
we call the vanilla Lie derivative $\mathfrak L_\xi$ acts on $e^a_{\;\,\mu}$ as if these are
$d$-many 1-forms that are mutually unrelated, so does not treat the
local Lorentz indices properly. This casual attitude extends to the spinor
indices, so that,
\bea
\mathfrak L_\xi\Psi=\xi^\mu\partial_\mu\Psi \ .
\eea
In physics literature, $\mathfrak L_\xi$ appears to be by far more prevalent, yet,
one can already see that turning off the Kosmann lift, $\hat\xi_K=0$,
cannot be achieved covariantly, as it equates a tensor,  $\hat\xi^{ab}_{V}$, to
a connection contracted with a vector, $\xi^\mu w_\mu^{\ ab}$.

Provided that we employ the Kosmann-lifted Lie derivative, $\mathscr L_\xi$,
we again find the identity $T=\hat\iT$ naturally emerging from the most general covariant Lagrangian
$\cL(\Psi,\mathscr D\Psi,\cdots)$, up to one-derivative on spinors but otherwise unrestricted.
Surprisingly, the form of the
energy-momentum tensor is even more robust than the Lagrangian in that the
addition of total derivative terms to $\cL$   is automatically screened out by
the  procedure we offer. As with the bosonic cases, we neither rely on the equation of
motion nor invoke any sort of ``improvement" for establishing the stated identity.

Finally, this brings us to the important question of how the Kosmann lift would figure into the computation
of the diffeomorphism anomalies. For this, we retrace the classic computation
by Alvarez-Gaume and Witten \cite{Alvarez-Gaume:1983ihn}, which turns out to have computed anomalies of
neither $\mathscr{L}_\xi$ nor $\mathfrak L_\xi$ even though they seemingly started out
with $\mathfrak L_\xi=\xi^\mu\partial_\mu$ as the generator on spinors. Nevertheless, starting with
$\mathscr L_\xi$ as the diffeomorphism generator instead, we arrive at the same old
anomaly polynomials and a modified extraction rule for the covariant anomalies.

We find that the only difference in the end due to the more sensible $\mathscr{L}_\xi$
is a matter of the simple factor 1/2 multiplying the covariant diffeomorphism anomaly
of the original computation. The same factor $1/2$ proves to be necessary to
put the consistent diffeomorphism anomaly on an equal footing with the more
familiar gauge counterpart, as we perform the anomaly descent. We believe
this very necessary numerical factor has been in effect employed on the consistent
side, without being properly recognized, for decades.

In a sense, the Kosmann lift completes this venerable computation,
teaching us that the diffeomorphism generator employed back then should be
more sensibly interpreted as the first of two covariant pieces in (\ref{KL_cov})
that constitutes $\mathscr L_\xi$. More importantly, our treatment puts the
derivation of the anomaly polynomials on solid ground and restores the
correct relationship between covariant anomaly and the commonly-quoted consistent
anomaly, which strangely enough was lacking in the existing literature.

\vskip 5mm
\noindent
N.B. The seed for this manuscript was developed for a graduate text \cite{V2} in preparation
by the senior author. We borrowed some relevant contents from this volume and condensed
in Sections 2 and 4, and in turn, the contents of Sections 5 and 6 here  are
to be imported to the said text.

\section{Noether Procedure Revisited}

\subsection{Noether Current and Gauge Current}

Let us warm up by recalling the Noether procedure for internal rotational symmetries.
The point of repeating this basic fact of life for any student of quantum
field theories will become apparent at the very end of the section, which
will be taken up to fix, once and for all, the more confusing story of the
Noether energy-momentum in the next section.

Consider the action
\bea
S(\phi)\;=\; \int d^{\,d}x\;\cL(\phi,\partial_\mu\phi)
\eea
that admits a global symmetry, i.e., an infinitesimal and position-independent
shift of $\phi$
\bea
\epsilon\boldsymbol\delta_{ \theta}\phi\; \equiv\;\boldsymbol\delta_{\epsilon \theta}\phi\; =\;\ii\epsilon\theta\phi \ ,
\eea
with  $\theta=\sum_C\theta^C t^C$ with Hermitian $t^C$'s that leaves the action invariant.

We then elevate $\theta$ to a position-dependent one and vary the action,
$\boldsymbol\delta_\theta\phi$, i.e., expanding the action in linear order
in $\epsilon\boldsymbol\delta_\theta$ then divide by $\epsilon$,
\bea
\int d^{\,d}x\;\boldsymbol\delta_\theta \cL(\phi,\partial_\mu\phi)&=& \int \sum_\phi\left(\frac{\boldsymbol\partial \cL}{\boldsymbol\partial\phi} \boldsymbol\delta_\theta \phi +
\frac{\boldsymbol\partial \cL}{\boldsymbol\partial(\partial_\mu\phi)} \partial_\mu (\boldsymbol\delta_\theta \phi)\right)\cr\cr
&=& \int \sum_\phi\left(\frac{\boldsymbol\partial \cL}{\boldsymbol\partial\phi} - \partial_\mu \frac{\boldsymbol\partial \cL}{\boldsymbol\partial(\partial_\mu\phi)}\right)\boldsymbol\delta_\theta \phi
-\int \partial_\mu \iJ^\mu_\theta\ ,
\eea
with
\bea
\iJ^\mu_\theta \equiv -\sum_\phi\frac{\boldsymbol\partial \cL}{\boldsymbol\partial(\partial_\mu\phi)}\boldsymbol\delta_\theta\phi\ .
\eea
{\it Turning off the position-dependence at the end} and using the equation of motion we arrive at
\bea
\partial_\mu \iJ^\mu_\theta = 0\ ,
\eea
which is the celebrated Noether's conservation law.

\subsubsection*{Gauge Currents}

Gauging such a symmetry means introducing a gauge connection $A=\sum_CA^Ct^C$,
\bea
\partial_\mu \phi\quad\rightarrow\quad D_\mu\phi\equiv (\partial_\mu-\ii A_\mu)\phi\ ,
\eea
whereby the action
\bea
S(\phi,A)\;=\; \int d^{\,d}x\;\cL(\phi,D_\mu\phi)
\eea
is invariant under spacetime dependent $\theta$,
\bea
S(\phi,A)=S(\phi+\epsilon\boldsymbol\delta_\theta\phi,A+\epsilon \boldsymbol\delta_\theta A) \ ,
\eea
with
\bea
\boldsymbol\delta_\theta\phi\;=\; \ii\theta \phi \ , \qquad \boldsymbol\delta_\theta A\;=\;d_A\theta =d\theta-\ii [A,\theta] \ .
\eea
This leads us to an alternate definition of the current,\footnote{We introduced the partial variation of the local functional,
\bea
\frac{\boldsymbol\delta L(f,\partial f,\cdots\,)}{\boldsymbol\delta f}\boldsymbol\delta f \;=\;\lim_{\epsilon\rightarrow 0} \frac{ L(f+\epsilon \, \boldsymbol\delta f,\partial f+\epsilon \, \boldsymbol\delta (\partial f),\cdots\,)- L(f,\partial f, \cdots\,)}{\epsilon} \ ,
\eea
to be used inside the integration over $x$, for which we will take care not
to integrate by parts freely. }
\bea\label{current}
J_C^\mu\equiv \frac{\boldsymbol\delta \cL}{\boldsymbol\delta A_\mu^C}\ ,
\eea
which is now  conserved in the covariant sense,
\bea
0=D_\mu J^\mu=\partial_\mu J^\mu -\ii (A_\mu J^\mu- J^\mu A_\mu) \ ,
\eea
with $J^\mu\equiv \sum_C J_C^\mu t^C $. Note that, for a minimally coupled scalar field,
\bea
%J_\theta^\mu =
\sum_C\theta^C J_C^\mu
=\sum_C\sum_\phi\frac{\boldsymbol\partial \cL}{\boldsymbol\partial(D_\mu\phi)}\frac{\theta^C\boldsymbol\partial(D_\mu\phi)}{\boldsymbol\partial A_\mu^C}=-\sum_\phi\frac{\boldsymbol\partial \cL(\phi,D_\mu\phi)}{\boldsymbol\partial(D_\mu\phi)}
\boldsymbol\delta_\theta\phi=\hat\iJ^\mu_\theta \ .
\eea
The right hand side is nothing but the covariantized version of
the Noether current. This of course reflects in part how we started the gauging process
by introducing a gauge field $A_\mu$ and contracting it against $\mathbbm{J}^\mu$.

All of the above elevate to curved spacetime almost verbatim, with the covariantized action,
\bea
S(\phi)\;=\; \int d^{\,d}x\,\sqrt{g}\;\cL(\phi,(\nabla_\mu- \ii A_\mu)\phi)\;=\; \int\cV\,\cL(\phi,(\nabla_\mu-\ii A_\mu)\phi)\ ,
\eea
where $\cV$ is the volume form. This covariantizes $\partial$ to $\nabla$, if $\phi$
is a more general tensor field. As long as $\boldsymbol\delta_\theta$ does not transform
the metric, nothing else changes. If one chooses to treat $\cV\cL$ itself as the $d$-form
Lagrangian density, the conservation law would translate to
\bea
0=d_A (J\lrcorner \,\cV )\ ,
\eea
where $d_A$ is the covariantized exterior derivative. Since this $(d-1)$-form
current naturally emerges when we start with  $\cV\cL$,
we will use the same notation $J$ in place of $J \lrcorner \, \cV$ as well
\bea
0=d_A  J\ ,
\eea
from now on. The context should make the distinction unambiguous.

\subsubsection*{How the Noether Current Equals the Gauge Current}

An instructive lesson can be learned by asking if and how an automatic agreement
between $\hat\iJ$ and $J$ occurs, once we gauge the action, forgetting for the
moment that the gauging itself started with $\iJ$. When we derive the Noether current $\iJ$ of internal symmetries,
we start out with
a position-dependent $\theta$  even though such $\theta$ transformation
does not preserve the action.
Instead, one says that this is a mere trick, as we will take constant $\theta$ at the
end of the day.
Nevertheless, this is a little odd thing to do when there is a perfectly sensible
gauged extension of the action that would be invariant under such position-dependent
$\theta$. Is there a better way to understand
the Noether procedure from the gauged version of the theory?

For this, we start from the gauged Lagrangian
$\cL(\phi, A)$ such that
\bea
 \boldsymbol\delta_\theta \cL(\phi, A) =0\ ,
\eea
identically for position-dependent $\theta$, where neither the
equation of motion nor an integration by parts is invoked.
Then we shall split this vanishing
net variation  into two mutually canceling parts. One is from the transformation
of $\phi$,
\bea\label{del_phi}
 \frac{\boldsymbol\delta \cL(\phi,A)}{\boldsymbol\delta\phi}\boldsymbol\delta_\theta\phi  \;=\;
  -\left(D_\mu\frac{\boldsymbol\partial \cL}{\boldsymbol\partial (D_\mu\phi)}\right) \boldsymbol\delta_\theta\phi -\partial_\mu \hat\iJ^\mu_\theta \ ,
\eea
 where $\hat\iJ_\theta$ is the covariantized version of
the Noether conservation law. The other, from $\boldsymbol\delta_\theta A$,   results in
\bea\label{del_A}
\frac{\boldsymbol\delta \cL(\phi,A)}{\boldsymbol\delta A}\boldsymbol\delta_\theta A  &=& {\rm tr}(\boldsymbol\delta_\theta A_\mu  J^\mu)
\cr\cr&=& -{\rm tr}\left(\theta \,(d_A J)\right)+  d\left({\rm tr}(\theta J)\right) \ ,
\eea
also to the linear order.

The  two combine to complete $\boldsymbol\delta_\theta$ of the action and must
cancel out identically,
\bea
0=\boldsymbol\delta_\theta\cL(\phi,A)
=\frac{\boldsymbol\delta \cL(\phi,A)}{\boldsymbol\delta\phi}\boldsymbol\delta_\theta\phi
+ \frac{\boldsymbol\delta \cL(\phi,A)}{\boldsymbol\delta A}\boldsymbol\delta_\theta A \ ,
\eea
even when $\theta$ is position-dependent, since this is precisely
what we mean by gauging an internal symmetry. For a scalar field
minimally coupled to $A$, it is not difficult to see by direct
computations that  the respective bulk terms in (\ref{del_phi})
and in (\ref{del_A}) cancel each other precisely.

As such, since position-dependent $\theta$ truly preserves the gauged action,
the two remaining total derivatives
must also cancel each other. This enforces ${\rm tr}(\theta J^\mu)=\hat\iJ_\theta^\mu$
prior to taking the divergences, for entirely arbitrary $\theta$, implying  $J=\hat\iJ$
where  ${\rm tr}(\theta \hat\iJ^\mu)\equiv \hat\iJ_\theta$.
It is important to emphasize how we neither invoked
the equation of motion nor threw away a total derivative term for this comparison.
The classical conservation law requires the equation of motion for either
current, but the equality of the two currents does not require one. The
equality proves to be an identity.

\subsection{Energy-Momentum}

The last observation on the identity between the Noether current
and the gauge current, $\hat\iJ=J$, is not something reflected in the typical text between
the Noether energy-momentum $\iT$ and the symmetric energy-momentum $T$
from the metric variation. Rather one often talks about
how $\iT$ for the spacetime translation, should be ``improved" toward $T$.
Here, we illustrate how such a perceived disparity may be rectified
by a simple generalization of our observation in the previous section.

Let us consider a matter action coupled to a curved spacetime by
appropriately elevating the derivatives to the covariant ones,
\bea
\int \cV\,\cL(\; \cdots\, ;g) \ ,
\eea
with the volume form $\cV$.
With such a minimal  coupling to the general metric, we immediately find
the energy-momentum tensor,
defined from the variation of the inverse metric,
\bea
T_{\mu\nu}\;\equiv\; -\frac{2}{\sqrt{g}}\frac{\delta}{\delta g^{\mu\nu}}\int \cV\,\cL\ ,
\eea
similar to the gauge current Eq.~(\ref{current}). This is clearly analogous
to the gauge currents $J$ of internal symmetries.

In the simplest example of a real scalar
\bea
&&\cL(\phi,\nabla_\mu\phi;g)\;=\;-\frac12g^{\mu\nu}\nabla_\mu \phi\nabla_\nu\phi -V(\phi)\cr\cr
& \Rightarrow& T_{\mu\nu}\;=\;\nabla_\mu\phi\nabla_\nu\phi-\frac12 g_{\mu\nu}(\nabla\phi)^2- g_{\mu\nu}V(\phi)\ ,
\eea
while for Maxwell theory we find
\bea
\cL(F;g)=-\frac{1}{4}F_{\mu\nu}F^{\mu\nu}
\quad \Rightarrow \quad
T_{\mu\nu}=F_{\mu\lambda}F_\nu^{\;\;\lambda} - \frac14 g_{\mu\nu}F^2\ ,
\eea
which are all conserved, upon the equation of motion. How do they stack up against
the Noether energy-momentum?

\subsubsection*{Noether Energy-Momentum, or Not}

In the flat spacetime,  the time translation and the
spatial translations are isometries, so the Noether procedure should generate
$d$-many conserved currents, or collectively a tensor with two spacetime indices.
We will denote  the resulting Noether energy-momentum tensor as $\iT^{\mu}_{\;\;\alpha}$.
For a scalar field $\phi$ and how $\cL(\phi,\partial_\mu\phi)$
is nominally affected by $\phi(x)\rightarrow \phi(x+\epsilon\xi)$, divided by
the infinitesimal  $\epsilon$,
\bea\label{Noether_Weinberg}
\boldsymbol\delta_\xi \cL&=&\frac{\boldsymbol\partial\cL}{\boldsymbol\partial\phi}\xi^\alpha\partial_\alpha \phi+ \frac{\boldsymbol\partial\cL}{\boldsymbol\partial(\partial_\mu\phi)} \partial_\mu(\xi^\alpha\partial_\alpha\phi)\cr\cr
&=&\xi^\alpha\left(\frac{\boldsymbol\partial\cL}{\boldsymbol\partial\phi}\partial_\alpha \phi + \frac{\boldsymbol\partial\cL}{\boldsymbol\partial(\partial_\mu\phi)}
(\partial_\mu\partial_\alpha\phi)\right)+ (\partial_\mu\xi^\alpha)\frac{\boldsymbol\partial\cL}{\boldsymbol\partial(\partial_\mu\phi)}
\partial_\alpha\phi\cr\cr
&=&\xi^\alpha\partial_\alpha \cL+(\partial_\mu\xi^\alpha) \frac{\boldsymbol\partial\cL}{\boldsymbol\partial(\partial_\mu\phi)}
\partial_\alpha\phi
%\cr\cr &\rightarrow& (\partial_\mu\xi^\alpha)\left( \frac{\boldsymbol\partial\cL}{\boldsymbol\partial(\partial_\mu\phi)}\partial_\alpha\phi -\delta^\mu_{\;\;\alpha}\cL\right)
\ .
\eea
Since the equation of motion extremizes the action for an
arbitrary variation subject to a boundary condition, this
variation should also vanish on shell.

Integrating the second piece by parts, we find
\bea\label{Noether_Weinberg_2}
0=\int d^dx\; \boldsymbol\delta_\xi \cL  \qquad\rightarrow \qquad 0=\int d^dx\; \xi^\alpha\,\partial_\mu\left( -\frac{\boldsymbol\partial\cL}{\boldsymbol\partial(\partial_\mu\phi)}
\partial_\alpha\phi + \delta^\mu_{\;\;\alpha}\cL\right)\ ,
\eea
leading us to the Noether energy-momentum,
\bea\label{scalar_TT}
\iT^\mu_{\;\;\alpha}\;\equiv\;-\frac{\boldsymbol\partial\cL}{\boldsymbol\partial(\partial_\mu\phi)}
\partial_\alpha\phi +\delta^\mu_{\;\;\alpha}\cL \ ,
\eea
with its conservation law,
\bea
\partial_\mu\iT^\mu_{\;\;\alpha}\;=\;0\ .
\eea
For scalars, it is easy to see that covariantized version $\hat\iT_{\mu\nu}$ of
$\iT_{\mu\nu}$, obtained from replacing $\partial$ by $\nabla$ and judicious insertions
of the metric,  is precisely equal to $T_{\mu\nu}$.

When it comes to the energy-momentum, there are more than one Noether procedure known.
The one here, borrowed from the venerable text by S. Weinberg, deviates from that of the internal symmetries
earlier. For instance,  we dropped a total derivative term here while for the internal
symmetries the conservation law itself came about from a total derivative
term. In particular, note how we chose not to transform the integration measure
$d^dx$, contrary to some early texts on the matter. Nevertheless, we
arrive at the same conventional expression for the energy-momentum tensor of
a scalar field. In fact, there is a very important reasoning
behind this choice as we will turn to later.

How should this generalize to fields with spin content? For the Maxwell field with
\bea
\cL= -\frac14 F_{\mu\nu}F^{\mu\nu}\ ,
\eea
a blind implementation of the above, say, $A_\mu(x)\rightarrow A_\mu(x+\epsilon \xi)$ would give
the Noether energy-momentum tensor of the form
\bea\label{T_wrong}
F^{\mu\lambda}\partial_\alpha A_\lambda - \frac14 \delta^\mu_{\;\;\alpha}F^2\ ,
\eea
which famously differs from the flat limit of $T^\mu_{\;\;\alpha}$ and, worse, is not even gauge-invariant.

There are various remedies that remove this discrepancy, such as adding ``improvement" \cite{Belinfante:1940,Rosenfeld:1940}
term $-\partial_\lambda( F^{\mu\lambda}A_\alpha)$ which is automatically divergence-free.
Another well-known approach for correcting this oddity is to demand the ordinary gauge
invariance along the middle steps, but this solution would be tailor-made for gauge theories.
The real problem is how the blind Noether procedure that brought us to (\ref{T_wrong})
is not natural, to begin with, given how it ignores the spin content of $A_\mu$. This
simple fact gives us a different, completely universal solution to this general quandary.

\subsection{$\hat{\iT}$ Must Always Equal $T$ }

As we hinted at the end of the gauge current discussion in the previous section,
much of such ambiguity about the vanilla Noether procedure originates from how we seemingly
perform a position-dependent ``symmetry" operation even though the latter does not
preserve the action. This intermediate procedure is considered a trick in the usual
Noether argument, instead, to be justified by removing the position-dependence
in the end. However, since we would integrate by parts along the way,
this leaves a logical possibility that the ambiguous middle step can
lead to ambiguity of the form of the Noether current thus obtained.

On the other hand, after the proper gauging procedure, we have an unambiguous form of $\boldsymbol\delta_\theta\phi$ that together with $\boldsymbol\delta_\theta A$ preserves the gauged action.
This means that when  we elevate the global symmetry by coupling to gauge fields, external
or dynamical, the potential ambiguity of the Noether procedure is resolved.
The same principle should apply to the general covariance, i.e., the translations elevated to much
bigger coordinate redundancy by coupling to the metric. We find some details that differ from the
above internal symmetry example, however.

For the symmetric energy-momentum tensor $T$, the role of $\boldsymbol\delta_\theta A$
is taken up by $\boldsymbol\delta_\xi g=\mathfrak L_\xi g$, i.e., by the Lie derivative
since, in curved spacetime, the ``gauged" translation  is nothing but the general
coordinate transformation. The symmetric energy-momentum $T$ follows from
varying the metric inside the Lagrangian as
\bea
\frac{\boldsymbol\delta\left( \cV\,\cL(\phi;g)\right)}{\boldsymbol\delta g^{\mu\alpha}}\boldsymbol\delta_\xi g^{\mu\alpha}
= \cV\,(\nabla^\mu\xi^\alpha)\, T(\phi;g)_{\mu\alpha}\ ,
\eea
whose integration by parts produces the divergence of $T$ which in turn vanishes on shell
because the Einstein tensor sitting on the other side of the $g$-equation of motion is
divergence-free as a mathematical identity.

This means that the variation of matter fields is not ambiguous but should be performed also
by the Lie derivative $\mathfrak L$, if the entire matter Lagrangian, now gauged, is to be
inert under the position-dependent transformation.
Revisiting the case of scalar fields,
the other transformation gives, with $\boldsymbol\delta_\xi \phi=\xi^\mu\partial_\mu\phi=\xi^\mu\nabla_\mu\phi$,
\bea\label{cov_noether_em}
\cV\; \frac{\boldsymbol\delta \cL(\phi;g)}{\boldsymbol\delta\phi}\boldsymbol\delta_\xi\phi
&=&\cV\,\left(\xi^\alpha\nabla_\alpha \cL(\phi;g)+(\nabla_\mu\xi^\alpha) \frac{\boldsymbol\partial\cL(\phi;g)}{\boldsymbol\partial(\nabla_\mu\phi)}
\nabla_\alpha\phi\right)\cr\cr
&=& d\left(\xi\lrcorner\cV \cL(\phi;g)\right)-  \cV\,(\nabla_\mu\xi^\alpha) \; \hat{\iT}(\phi;g)^\mu_{\;\;\alpha} \ ,
\eea
with the covariantized $\hat{\iT}$ of the Noether energy-momentum
${\iT}$ we have computed in (\ref{scalar_TT}).

Unlike the internal symmetries, however, we have
\bea\label{delta_xi}
\boldsymbol\delta_\xi (\cV\cL) = \mathfrak L_\xi(\cV\cL)=d(\xi\lrcorner \cV\cL)\ ,
\eea
instead of vanishing identically.
Starting from this universal observation, we may  split the left hand side
into two parts, one from $\boldsymbol\delta_\xi g$ generating $T$
while the other from $\boldsymbol\delta_\xi \phi$ generating a covariant version $\hat{\iT}$ of $\iT$.
This universal fact (\ref{delta_xi}) then implies that these two combined should produce
\bea
 \frac{\boldsymbol\delta\left( \cV\,\cL(\phi;g)\right)}{\boldsymbol\delta g^{\mu\alpha}}\boldsymbol\delta_\xi g^{\mu\alpha}
 +\cV\; \frac{\boldsymbol\delta \cL(\phi;g)}{\boldsymbol\delta\phi}\boldsymbol\delta_\xi\phi =
 d(\xi\lrcorner \cV\cL(\phi, g))\ ,
\eea
which is possible only if
\bea
T(\phi;g)_{\mu\alpha} =\hat{\iT}(\phi;g)_{\mu\alpha} \ .
\eea
This agreement  for a scalar theory has been seen from explicit
computations earlier, but the line of thought here suggests that
the same should happen for any type of matter field as long as it is
covariantly coupled to the metric.

As with the internal gauge symmetry example above, we have invoked neither
an equation of motion nor an integration by parts in justifying the identity here,
although for individual conservation laws one needs those.
A hint for the above line of thoughts we followed to fix the Noether procedure once and for all
is found in how, in (\ref{Noether_Weinberg}) and (\ref{Noether_Weinberg_2}),
we did not transform the integration measure $d^d x$ under $\boldsymbol\delta_\xi$.
The measure is   a special case of the volume form $\cV$ which is in turn defined by $g$, once we gauge the action.
As such, we must not transform it for the Noether side in general curved spacetimes,
so the same should hold in the flat spacetime as well.

The key ingredient in the above reasoning is that  we must
use $\boldsymbol\delta_\xi\rightarrow \mathfrak L_\xi$
for the position-dependent ``translation" for the Noether procedure.
For the Maxwell field, we therefore use
\bea
\boldsymbol\delta_\xi F_{\alpha\beta} \;=\;
\mathfrak L_\xi F_{\alpha\beta}
%&=& \partial_\alpha (\xi^\mu F_{\mu\beta})- \partial_\beta(\xi^\mu F_{\mu\alpha})\cr\cr
%&=& (\partial_\alpha \xi^\mu) F_{\mu\beta}- (\partial_\beta\xi^\mu) F_{\mu\alpha}-\xi^\mu(\partial_\alpha F_{\beta\mu} +\partial_\beta F_{\mu\alpha})\cr\cr
&=&\xi^\mu\nabla_\mu F_{\alpha\beta}+ (\nabla_\alpha \xi^\mu) F_{\mu\beta}+ (\nabla_\beta\xi^\mu) F_{\alpha\mu}\ ,
\eea
for the Noether side on properly covariantized action, with the help of
$d\mathfrak L_\xi =\mathfrak L_\xi d$. The same would happen with Yang-Mills case as well.
Starting with the usual Maxwell action $\cL(F;g)=- F^2/4$, we find
\bea\label{Noether_Maxwell}
&&\cV\, \frac{\boldsymbol\delta \cL}{\boldsymbol\delta A} \,\boldsymbol\delta_\xi A
\;=\; \cV\,\frac{\boldsymbol\partial \cL}{\boldsymbol\partial F} \,\boldsymbol\delta_\xi F \cr\cr
%&=& -\frac12 \int\sqrt{g}\;g^{\gamma\alpha}g^{\delta\beta} F_{\gamma\delta}\boldsymbol\delta_\xi F_{\alpha\beta}\cr\cr
&=&
-\frac12 \,\cV\, g^{\gamma\alpha}g^{\delta\beta}
F_{\gamma\delta}\left((\nabla_\alpha \xi^\mu)
F_{\mu\beta}+ (\nabla_\beta\xi^\mu) F_{\alpha\mu}
+\xi^\mu\nabla_\mu F_{\alpha\beta} \right)\ ,
\eea
where we performed the variation of $F$ but kept $g$ untouched.
We can isolate the Noether current from this in two different manners.

The more conventional route
is to integrate by parts the first two terms of  (\ref{Noether_Maxwell})  and throw away
total derivative terms to obtain
\bea
\int \cV\;\frac{\boldsymbol\delta \cL}{\boldsymbol\delta A} \boldsymbol\delta_\xi A&\rightarrow&\int \sqrt{g}\;\xi^\mu\nabla^\alpha\left( F_{\alpha}^{\;\;\beta}F_{\mu\beta}-\frac14 g_{\alpha\mu}F^2 \right)\ ,
%\;=\; \int \sqrt{g}\;\xi^\mu\nabla^\alpha \hat{\iT}_{\alpha\mu}\ .
\eea
leading to the covariantized Noether energy-momentum of the form,
\bea
\hat{{\iT}}_{\alpha\mu}\equiv F_\alpha^{\;\;\beta}F_{\mu\beta}-\frac14 g_{\alpha\mu}F^2 \ .
\eea
We already see that the form of $\hat\iT$  equals the symmetric energy-momentum $T$.

Alternatively, we may rewrite the last term of (\ref{Noether_Maxwell}), as if we are
performing an integration by parts, instead
to find
\bea\label{Noether_EM_2}
 \cV\;\frac{\boldsymbol\delta \cL}{\boldsymbol\delta A} \boldsymbol\delta_\xi A\;=\; d\left(\xi\lrcorner\cV \cL\right)- \cV\,(\nabla_\mu\xi^\alpha) \; \hat{\iT}^\mu_{\;\;\alpha} \ ,
\eea
which, combined with the universal fact (\ref{delta_xi}), produces
\bea
\cV\;\frac{\boldsymbol\delta \cL}{\boldsymbol\delta A} \boldsymbol\delta_\xi A+
 \frac{\boldsymbol\delta(\cV\cL)}{\boldsymbol\delta g^{\mu\alpha}}\boldsymbol\delta_\xi g^{\mu\alpha}
%\;=\;\int \boldsymbol\delta_\xi (\cV\cL)
\;=\; d\left(\xi\lrcorner\cV \cL\right)\ ,
\eea
as advertised. Once the metric is introduced,
how one performs $\boldsymbol\delta_\xi$ for fields on general manifolds
should not be really a matter of choice, so the perceived ambiguity of
$\iT$ for the Maxwell theory is pretty much an artefact of
ill-conceived transformation rules. As with the gauge current example, the equality
$\hat\iT=T$ does not require the equation of motion or dropping a total derivative,
even though the classical conservation law would need such steps.

Even with the flat spacetime, in retrospect, the Lie derivative is unavoidable
if we started with curvilinear coordinates. Except in the Cartesian coordinates,
no one can claim that the shift $\xi^\mu$ is constant, simply because curvilinear
$x^\mu$'s behave in a very complicated manner even under the simple translational isometry. To
carry out the right symmetry operation, the Lie derivative enters in an essential
manner. An operation like $A_\mu(x^\alpha)\rightarrow A_\mu(x^\alpha+\epsilon\xi^\alpha)$
becomes nonsensical even in the Minkowski spacetime, when we regard $A_\mu$ component-wise.
How a position shift affects the fields should be independent of such coordinate choices
and the only such operation available, curved or not, would lead to the Lie derivative.

\section{Tensor Fields in General}

The key idea behind $\hat\iT=T$ outlined in the previous section, i.e., how the variation
of the metric and the variation of the matter fields must cancel each other neatly for
a general covariant Lagrangian leaving behind a universal total derivative, is such a natural one.
In particular, its execution for a common energy-momentum tensor for the Maxwell theory
is not new \cite{Saravi:2002}. The same general thought
should be applicable to any matter field  coupled to the
metric covariantly, yielding $\hat\iT=T$, yet the procedure for scalars and gauge fields shown in the previous
section works verbatim when the matter
Lagrangian does not involve the connection explicitly.

When the connection enters $\cL$, i.e., when the covariant derivative rather than the partial derivative
is needed in the matter action as well, $T$ comes about only after a partial integration since
the variation of the Levi-Civita connection would be written as covariant derivatives acting
on the varied metric components. Only if something similar happens for $\hat\iT$ and only if the
respective total derivatives cancel each other out identically, the idea outlined above would
enforce $\hat{\iT}=T$ literally. Crucially, all of these should occur before we remove the
derivative in $\nabla^{(\mu}\xi^{\nu)}$ by an integration by parts.

Here we will show that this is indeed the case for Lagrangians that involve arbitrary tensor
fields with up to one derivative on the matter fields. Although we
dote on  a tensor field of type $\Phi\indices{_\beta^\gamma}$ with one covariant index
and one contravariant one, the analysis extends straightforwardly for tensors with more coordinate
indices.

\subsection{Symmetric Energy-Momentum}

We shall consider the covariantized action with at most one derivative on the field,
i.e., in the form
\bea
\cL(\Phi\indices{_\beta^\gamma},\nabla_\lambda\Phi\indices{_\beta^\gamma};g^{\mu\nu},g_{\rho\sigma}) \ .
\eea
The symmetric energy-momentum follows from the variation of the matter action with respect to $g^{\mu\nu}$, i.e.,
\bea
\frac{\boldsymbol\delta(\cV\cL)}{\boldsymbol\delta g^{\mu\nu}}\boldsymbol\delta g^{\mu\nu}
\;=\;\cV
\left(
-\frac{1}{2}g_{\mu\nu}\cL
+\frac{\boldsymbol\delta\cL}{\boldsymbol\delta g^{\mu\nu}}
\right)
\boldsymbol\delta g^{\mu\nu} \ ,
\eea
and the second term  includes pieces that come from the variation of the connection,
\bea
%\frac{\boldsymbol\delta\cL}{\boldsymbol\delta g^{\mu\nu}}\boldsymbol\delta g^{\mu\nu}
%\!&=&\!\frac{\boldsymbol\partial\cL}{\boldsymbol\partial g^{\mu\nu}}\boldsymbol\delta g^{\mu\nu}
%+\frac{\boldsymbol\partial\cL}{\boldsymbol\partial g_{\rho\sigma}}\boldsymbol\delta g_{\rho\sigma}
%+\frac{\boldsymbol\partial\cL}{\boldsymbol\partial(\nabla_\lambda\Phi\indices{_\beta^\gamma})}
%(\boldsymbol\delta\nabla)_\lambda\Phi\indices{_\beta^\gamma}
%\cr\cr
%\!&=&\!
%\left(
%\frac{\boldsymbol\partial\cL}{\boldsymbol\partial g^{\mu\nu}}
%-\frac{\boldsymbol\partial\cL}{\boldsymbol\partial g_{\rho\sigma}}g_{\rho\mu}g_{\sigma\nu}
%\right)
%\boldsymbol\delta g^{\mu\nu}
%\cr\cr &&
-\,\boldsymbol\delta\Gamma\indices{^\gamma_{\lambda\beta}}
\left(
\frac{\boldsymbol\partial\cL}{\boldsymbol\partial(\nabla_\lambda\Phi\indices{_\beta^\sigma})}
\Phi\indices{_\gamma^\sigma}
-\frac{\boldsymbol\partial\cL}{\boldsymbol\partial(\nabla_\lambda\Phi\indices{_\sigma^\gamma})}
\Phi\indices{_\sigma^\beta}
\right) \ ,
\eea
which will incur additional steps absent in the scalar and the gauge field cases.

For the latter types of terms, a tensor $\cC$
\bea\label{cC}
(\cC^{\rho\mu\nu})_{\lambda\alpha\beta}
\;\equiv\;
\delta\indices{^\rho_\lambda}\delta\indices{^\mu_\alpha}\delta\indices{^\nu_\beta}
+\delta\indices{^\rho_\beta}\delta\indices{^\mu_\lambda}\delta\indices{^\nu_\alpha}
-\delta\indices{^\rho_\alpha}\delta\indices{^\mu_\beta}\delta\indices{^\nu_\lambda} \
\eea
comes in handy, as it allows us to write
\bea
\boldsymbol\delta\Gamma\indices{^\gamma_{\lambda\beta}}
%&=&-\frac{1}{2} \left(
%\delta\indices{^\rho_\lambda}\delta\indices{^\gamma_{\!\!(\mu}}g_{\nu)\beta}^{}
%+\delta\indices{^\rho_\beta}g_{\lambda(\mu}^{}\delta\indices{^\gamma_{\nu)}}
%-g^{\rho\gamma}g_{\beta(\mu}g_{\nu)\lambda}
%\right)
%\nabla_\rho\boldsymbol\delta g^{\mu\nu}
%\cr\cr
&=&-\frac{1}{2}(\cC\indices{^\rho_{\!\!(\mu\nu)}}\!)\indices{_\lambda^\gamma_\beta}
\nabla_\rho\boldsymbol\delta g^{\mu\nu}
\ .
\eea
The same $\cC$ will make an appearance later for fermions as well. Here, we end up with
\begin{align}\label{metric_variation_tensor}
&\frac{\boldsymbol\delta(\cV\cL)}{\boldsymbol\delta g^{\mu\nu}}\boldsymbol\delta g^{\mu\nu}
\nonumber\\[0.5ex]
&\!\!
=\cV
\left[
-\frac{1}{2}T_{\mu\nu}\boldsymbol\delta g^{\mu\nu}
\!+\!\frac{1}{2}
\nabla_\rho\!
\left[(\cC\indices{^\rho_{\!\!(\mu\nu)}}\!)\indices{_\lambda^\gamma_\beta}
\boldsymbol\delta g^{\mu\nu}\!
\left(
\frac{\boldsymbol\partial\cL}{\boldsymbol\partial(\nabla_\lambda\Phi\indices{_\beta^\sigma})}
\Phi\indices{_\gamma^\sigma}
\!-\!\frac{\boldsymbol\partial\cL}{\boldsymbol\partial(\nabla_\lambda\Phi\indices{_\sigma^\gamma})}
\Phi\indices{_\sigma^\beta}
\right)
\right]
\right] ,
\end{align}
where the symmetric energy-momentum is explicitly given as
\bea
T_{\mu\nu}
\!&=&\!g_{\mu\nu}\cL
-2\frac{\boldsymbol\partial\cL}{\boldsymbol\partial g^{\mu\nu}}
+2\frac{\boldsymbol\partial\cL}{\boldsymbol\partial g_{\rho\sigma}}g_{\rho\mu}g_{\sigma\nu}
\cr\cr
\!&&\!
+(\cC\indices{^\rho_{\!\!(\mu\nu)}}\!)\indices{_\lambda^\gamma_\beta}\nabla_\rho\!
\left(
\frac{\boldsymbol\partial\cL}{\boldsymbol\partial(\nabla_\lambda\Phi\indices{_\beta^\sigma})}
\Phi\indices{_\gamma^\sigma}
-\frac{\boldsymbol\partial\cL}{\boldsymbol\partial(\nabla_\lambda\Phi\indices{_\sigma^\gamma})}
\Phi\indices{_\sigma^\beta}
\right) \ ,
\eea
in the first term with ${\boldsymbol\delta g_{\rho\sigma}}=-{\boldsymbol\delta g^{\mu\nu}} g_{\mu\rho} g_{\sigma\nu}$.
Note how the other piece in (\ref{metric_variation_tensor}), also involving $\cC$, is a total derivative.
As such,  $T_{\mu\nu}$ here is the symmetric energy-momentum tensor that enters the Einstein equation.

Under the infinitesimal diffeomorphism $\boldsymbol\delta_\xi g^{\mu\nu}=\mathfrak L_\xi g^{\mu\nu}=-2\nabla^{(\mu}\xi^{\nu)}$, the variation becomes
\bea\label{Symmetric_Mixed}
\frac{\boldsymbol\delta(\cV\cL)}{\boldsymbol\delta g^{\mu\nu}}\boldsymbol\delta_\xi g^{\mu\nu}
=\cV \left[
(\nabla^\mu\xi^\nu)T_{\mu\nu} + \nabla_\rho \,\cS^\rho\right]\ ,
\eea
with
\bea
\cS^\rho\equiv -
(\cC\indices{^\rho_{\!\!(\mu\nu)}}\!)\indices{_\lambda^\gamma_\beta}
(\nabla^\mu\xi^\nu)\!
\left(
\frac{\boldsymbol\partial\cL}{\boldsymbol\partial(\nabla_\lambda\Phi\indices{_\beta^\sigma})}
\Phi\indices{_\gamma^\sigma}
-\frac{\boldsymbol\partial\cL}{\boldsymbol\partial(\nabla_\lambda\Phi\indices{_\sigma^\gamma})}
\Phi\indices{_\sigma^\beta}
\right) .
\eea
Integrating by parts leads us to the conservation law $\nabla^\mu T_{\mu\nu}=0$ in the bulk,
again consistent with the Einstein equation. Generalization to arbitrary tensor fields is
straightforward. Also,  the reduction  to scalars and gauge fields,
devoid of  connection contributions, coincides with the previous section.

Below we will see that the Noether variation based on the Lie derivative of $\Phi$
produces an expression that in part should cancel away (\ref{Symmetric_Mixed}). This will
split into three parts. The universal part $d(\xi\lrcorner\,\cV\cL)$ comes about after some
manipulation while, among the remainder that is supposed to cancel  (\ref{Symmetric_Mixed}),
one piece has $-\hat\iT$ in place of $T$ and the other is $- \nabla_\rho \cS^\rho$.
The latter cancels the total derivative piece in (\ref{Symmetric_Mixed}), which
enforces the identity $T=\hat \iT$, again as anticipated, regardless of the details of
the Lagrangian.

\subsection{Noether Energy-momentum}

For a tensor field $\Phi\indices{_\beta^\gamma}$, the Noether procedure should be performed using
\bea
\boldsymbol\delta_\xi\Phi\indices{_\beta^\gamma}
&=&\mathfrak L_\xi\Phi\indices{_\beta^\gamma}\cr\cr
&=&\xi^\alpha\nabla_\alpha\Phi\indices{_\beta^\gamma}
+(\nabla_\beta\xi^\alpha)\Phi\indices{_\alpha^\gamma}
-(\nabla_\alpha\xi^\gamma)\Phi\indices{_\beta^\alpha}
\ .
\eea
With the Lagrangian in the form $\cL(\Phi,\nabla\Phi;g)$, we have
\bea
\!\!\!\!
\boldsymbol\delta_\xi\cL\,\biggr\vert_{g\;{\rm fixed}}
\!&=&\!\frac{\boldsymbol\partial\cL}{\boldsymbol\partial\Phi\indices{_\beta^\gamma}}\boldsymbol\delta_\xi\Phi\indices{_\beta^\gamma}
+\frac{\boldsymbol\partial\cL}{\boldsymbol\partial(\nabla_{\lambda}\Phi\indices{_\beta^\gamma})}\nabla_{\lambda}(\boldsymbol\delta_\xi\Phi\indices{_\beta^\gamma})
%\cr\cr
%\!&=&\!\xi^\alpha
%\left(
%\frac{\boldsymbol\partial\cL}{\boldsymbol\partial\Phi\indices{_\beta^\gamma}}\nabla_\alpha\Phi\indices{_\beta^\gamma}
%+\frac{\boldsymbol\partial\cL}{\boldsymbol\partial(\nabla_\lambda\Phi\indices{_\beta^\gamma})}\nabla_\alpha(\nabla_\lambda\Phi\indices{_\beta^\gamma})
%\right)
%\cr\cr
%\!&&\!+\nabla_\mu\xi^\alpha
%\left(
%\frac{\boldsymbol\partial\cL}{\boldsymbol\partial(\nabla_\mu\Phi\indices{_\beta^\gamma})}\nabla_\alpha\Phi\indices{_\beta^\gamma}
%\right.
%\cr\cr
%\!&&\!+
%\left.
%\frac{\boldsymbol\partial\cL}{\boldsymbol\partial\Phi\indices{_\mu^\gamma}}\Phi\indices{_\alpha^\gamma}
%+\frac{\boldsymbol\partial\cL}{\boldsymbol\partial(\nabla_\lambda\Phi\indices{_\mu^\gamma})}\nabla_\lambda\Phi\indices{_\alpha^\gamma}
%-\frac{\boldsymbol\partial\cL}{\boldsymbol\partial\Phi\indices{_\beta^\alpha}}\Phi\indices{_\beta^\mu}
%-\frac{\boldsymbol\partial\cL}{\boldsymbol\partial(\nabla_\lambda\Phi\indices{_\beta^\alpha})}\nabla_\lambda\Phi\indices{_\beta^\mu}
%\right)
%\cr\cr
%\!&&\!+\frac{\boldsymbol\partial\cL}{\boldsymbol\partial(\nabla_\lambda\Phi\indices{_\beta^\gamma})}
%\left(
%(\nabla_\lambda\nabla_\beta\xi^\alpha)\Phi\indices{_\alpha^\gamma}
%-(\nabla_\lambda\nabla_\alpha\xi^\gamma)\Phi\indices{_\beta^\alpha}
%+\xi^\alpha[\nabla_\lambda,\nabla_\alpha]\Phi\indices{_\beta^\gamma}
%\right)
%
\ ,
\eea
which can be organized into
\bea
\!\!\!\!
\boldsymbol\delta_\xi\cL\,\biggr\vert_{g\;{\rm fixed}}
\!&=&\!\xi^\alpha\nabla_\alpha\cL
+\nabla_\mu\xi^\alpha
\left(
\frac{\boldsymbol\partial\cL}{\boldsymbol\partial(\nabla_\mu\Phi\indices{_\beta^\gamma})}\nabla_\alpha\Phi\indices{_\beta^\gamma}
\right.
\cr\cr
\!&&\!+
\left.
\frac{\boldsymbol\partial\cL}{\boldsymbol\partial\Phi\indices{_\mu^\gamma}}\Phi\indices{_\alpha^\gamma}
+\frac{\boldsymbol\partial\cL}{\boldsymbol\partial(\nabla_\lambda\Phi\indices{_\mu^\gamma})}\nabla_\lambda\Phi\indices{_\alpha^\gamma}
-\frac{\boldsymbol\partial\cL}{\boldsymbol\partial\Phi\indices{_\beta^\alpha}}\Phi\indices{_\beta^\mu}
-\frac{\boldsymbol\partial\cL}{\boldsymbol\partial(\nabla_\lambda\Phi\indices{_\beta^\alpha})}\nabla_\lambda\Phi\indices{_\beta^\mu}
\right)
\cr\cr
\!&&\!+
\left(
\nabla_\lambda\nabla_\beta\xi^\gamma
+[\nabla_\beta,\nabla^\gamma]\xi_\lambda
\right)
\left(
\frac{\boldsymbol\partial\cL}{\boldsymbol\partial(\nabla_\lambda\Phi\indices{_\beta^\sigma})}\Phi\indices{_\gamma^\sigma}
-\frac{\boldsymbol\partial\cL}{\boldsymbol\partial(\nabla_\lambda\Phi\indices{_\sigma^\gamma})}\Phi\indices{_\sigma^\beta}
\right)
\ ,
\eea
after some hefty algebra.

We used the usual gymnastics on the covariant derivative, such as
\bea
\xi^\alpha[\nabla_\lambda,\nabla_\alpha]\Phi\indices{_\beta^\gamma}
&=&\xi^\alpha
\left(
-R\indices{^\mu_{\beta\lambda\alpha}}\Phi\indices{_\mu^\gamma}
+R\indices{^\gamma_{\mu\lambda\alpha}}\Phi\indices{_\beta^\mu}
\right)
\cr\cr
%&=&
%R\indices{_{\lambda\alpha\beta}^\mu}\xi^\alpha\cdot\Phi\indices{_\mu^\gamma}
%-R\indices{_{\lambda\alpha\mu}^\gamma}\xi^\alpha\cdot\Phi\indices{_\beta^\mu}
%\cr\cr
&=&
([\nabla_\beta,\nabla^\mu]\xi_\lambda)\Phi\indices{_\mu^\gamma}
-([\nabla_\mu,\nabla^\gamma]\xi_\lambda)\Phi\indices{_\beta^\mu}
\ .
\eea
Another such is
\bea
\!\!\!\!\!\!\!\!\!\!
\nabla_\lambda\nabla_\beta\xi^\gamma
+[\nabla_\beta,\nabla^\gamma]\xi_\lambda
%\!&=&\!\nabla_\rho\nabla_\mu\xi_\nu
%\left(
%\delta\indices{^\rho_\lambda}g^{\gamma\nu}\delta\indices{^\mu_\beta}
%+\delta\indices{^\rho_\beta}\delta\indices{^\nu_\lambda}g^{\mu\gamma}
%-g^{\rho\gamma}\delta\indices{^\mu_\beta}\delta\indices{^\nu_\lambda}
%\right)
%\cr\cr
%\!&=&\!\nabla_\rho\nabla_\mu\xi_\nu
%\left(
%\delta\indices{^\rho_\lambda}g^{\gamma(\nu}\delta\indices{^{\mu)}_{\!\!\beta}}
%+\delta\indices{^\rho_\beta}\delta\indices{^{(\nu}_{\lambda}}g^{\mu)\gamma}
%-g^{\rho\gamma}\delta\indices{^{(\mu}_\beta}\delta\indices{^{\nu)}_{\!\!\lambda}}
%\right)
%\cr\cr
\!&=&\!
\nabla_\rho\nabla_\mu\xi_\nu
\,(\cC^{\rho(\mu\nu)})\indices{_{\lambda}^\gamma_\beta}
\eea
that comes about, thanks to
\bea
\nabla_\rho\nabla_\mu\xi_\nu \,(\cC^{\rho[\mu\nu]})\indices{_{\lambda}^\gamma_\beta}
&=&\frac{1}{2}
\left(
[\nabla_\lambda,\nabla_\beta]\xi^\gamma+[\nabla_\beta,\nabla^\gamma]\xi_\lambda+[\nabla^\gamma,\nabla_\lambda]\xi_\beta
\right)
\cr\cr
&=&\frac{1}{2}
\left(
R\indices{_{\lambda\beta}^\gamma_\alpha}+R\indices{_\beta^\gamma_{\lambda\alpha}}+R\indices{^\gamma_{\lambda\beta\alpha}}
\right)
\xi^\alpha\;=\;0 \ ,
\eea
by virtue of the combinatoric symmetry of the Riemann tensor.

Eventually, all of these lead to
%\bea
%\!\!\!\!\! \!\!\!\!\! \!\!\!\!\!
%\cV\frac{\boldsymbol\delta\cL}{\boldsymbol\delta\Phi\indices{_\beta^\gamma}}\boldsymbol\delta_\xi\Phi\indices{_\beta^\gamma}
%\!&=&\!\nabla_\alpha(\xi^\alpha\cL)
%-(\nabla_\mu\xi^\alpha)\hat\iT\indices{^\mu_\alpha}
%\cr\cr
%\!&&\!+(\cC^{\rho(\mu\nu)})\indices{_{\lambda}^\gamma_\beta}
%\nabla_\rho\!
%\left[
%\nabla_\mu\xi_\nu\!
%\left(
%\frac{\boldsymbol\partial\cL}{\boldsymbol\partial(\nabla_\lambda\Phi\indices{_\beta^\sigma})}
%\Phi\indices{_\gamma^\sigma}
%\!-\!\frac{\boldsymbol\partial\cL}{\boldsymbol\partial(\nabla_\lambda\Phi\indices{_\sigma^\gamma})}
%\Phi\indices{_\sigma^\beta}
%\right)
%\right] ,
%\eea
\bea\label{Noether_Mixed}
\frac{\boldsymbol\delta(\cV\cL)}{\boldsymbol\delta\Phi\indices{_\beta^\gamma}}\boldsymbol\delta_\xi\Phi\indices{_\beta^\gamma}
=d(\xi\lrcorner\cV\cL)+
\cV\left[(\nabla_\mu\xi^\alpha)(-\hat\iT\indices{^\mu_\alpha})-\nabla_\rho\,\cS^\rho\right]\ .
\eea
Here, $\cS^\rho$ is identical to its namesake in (\ref{Symmetric_Mixed}) which happens
because the partial derivative with respect to the connection has a simple relation to
the derivative with respect to $\nabla\Phi$. The unique bulk term
defines the Noether energy-momentum $\hat\iT\indices{^\mu_\alpha}$ as follows,
\bea
\hat\iT\indices{^\mu_\alpha}
\!&=&\!\delta\indices{^\mu_\alpha}\cL
-\frac{\boldsymbol\partial\cL}{\boldsymbol\partial(\nabla_\mu\Phi\indices{_\beta^\gamma})}\nabla_\alpha\Phi\indices{_\beta^\gamma}
\cr\cr
\!&&\!-\frac{\boldsymbol\partial\cL}{\boldsymbol\partial\Phi\indices{_\mu^\gamma}}\Phi\indices{_\alpha^\gamma}
-\frac{\boldsymbol\partial\cL}{\boldsymbol\partial(\nabla_\lambda\Phi\indices{_\mu^\gamma})}\nabla_\lambda\Phi\indices{_\alpha^\gamma}
+\frac{\boldsymbol\partial\cL}{\boldsymbol\partial\Phi\indices{_\beta^\alpha}}\Phi\indices{_\beta^\mu}
+\frac{\boldsymbol\partial\cL}{\boldsymbol\partial(\nabla_\lambda\Phi\indices{_\beta^\alpha})}\nabla_\lambda\Phi\indices{_\beta^\mu}
\cr\cr
\!&&\!+(\cC^{\rho(\mu\nu)})\indices{_{\lambda}^\gamma_\beta}\,g_{\nu\alpha}\nabla_\rho\!
\left(
\frac{\boldsymbol\partial\cL}{\boldsymbol\partial(\nabla_\lambda\Phi\indices{_\beta^\sigma})}\Phi\indices{_\gamma^\sigma}
-\frac{\boldsymbol\partial\cL}{\boldsymbol\partial(\nabla_\lambda\Phi\indices{_\sigma^\gamma})}\Phi\indices{_\sigma^\beta}
\right) \ .
\eea

\subsection{Equality of $\hat\iT$ and  $T$}

Combining (\ref{Symmetric_Mixed}) and (\ref{Noether_Mixed}), we have
\bea
\boldsymbol\delta_\xi(\cV\cL)
&=&\frac{\boldsymbol\delta(\cV\cL)}{\boldsymbol\delta g^{\mu\nu}}\boldsymbol\delta_\xi g^{\mu\nu}
+\frac{\boldsymbol\delta(\cV\cL)}{\boldsymbol\delta\Phi\indices{_\beta^\gamma}}
\boldsymbol\delta_\xi\Phi\indices{_\beta^\gamma}
\cr\cr
&=&d(\xi\lrcorner\cV\cL)
+\cV\,(\nabla^\mu\xi^\nu)(T_{\mu\nu}-\hat\iT_{\mu\nu}) +\cV\left[ \cancel{\nabla_\rho\,\cS^\rho}-\cancel{\nabla_\rho\,\cS^\rho}\right]    \ ,
\eea
which, together with the universal property of the covariant Lagrangian density  $\boldsymbol\delta_\xi(\cV\cL)=  d(\xi\lrcorner\cV\cL)$,
gives the desired identity,
\bea
T_{\mu\nu}=\hat\iT_{\mu\nu}\ .
\eea
Note that we neither relied on the equation of motion nor threw away total derivative terms along the way.

\section{Vielbein, Spinors, and the Kosmann Lift}

Now that we have emphasized the role of the Lie derivative in the Noether procedure for the
energy-momentum tensor, we need to take a step back and consider how the Lie derivative should
act on objects with spacetime indices in the form of the local Lorentz index and the spinor index.

Most bosonic fields we encounter in physics are naturally tensors with coordinate indices. For example,
the gauge fields are locally 1-forms, or more precisely a connection 1-form with the coordinate
indices labeling its spacetime components naturally. The metric is a symmetric rank-two
covariant tensor with two coordinate indices. There are additional complications for
the Christoffel symbols, due to being connection for the Riemannian geometry, but one
can derive the action of the Lie derivative starting from the usual Lie derivative on the metric. For other
bosonic matter fields, the actions of Lie derivative are well understood.

A little more subtle is how the Lie derivative should act on geometrical objects with
the local Lorentz index or with the spinor index. This is a pretty acute issue
for spinors, in particular, since the Clifford algebra  is written naturally in the vielbein basis,
\bea
\{\gamma^a,\gamma^b\}=2\eta^{ab}\ .
\eea
One sometimes refers to the Dirac matrices
$\gamma^\mu$ in curved spacetime, but this is merely a shorthand notation for
\bea
\gamma^a e_a^{\;\,\mu}\ .
\eea
The usual statement about covariant constant $\gamma^\mu$'s traces back to how $\gamma^a$'s
are really constant matrices, modulo $SO$ rotations.

When the local Lorentz indices and the spinor indices are thus indispensable, one must ask
how the Lie derivative would handle objects equipped with these indices. In physics and mathematics
literature, one can find two  mutually conflicting treatments in this regard. In this intermediate
section, we will take a brief detour for these; in the next section we will end up advocating
what is known as the Kosmann lift for the correct version of the Lie derivative in our
current context of producing the right Noether energy-momentum tensor. The
rest of the section is, as noted earlier, borrowed and condensed from Ref.~\cite{V2}.

For the rest of this note, we take care to employ several distinct notations for the covariant derivatives,
in an incremental manner, depending on up to which bundles they take into account.
We start with $\nabla_\mu$ for (co-)tangent bundles and tensor products thereof,
which gives us the usual Levi-Civita connection expressed via the Christoffel symbols.
The notation $D_\mu$ is introduced when it becomes aware of the frame bundle, i.e., of the local Lorentz indices.
With spinors introduced, $\sD_\mu$ is the relevant covariant derivative in the
absence of gauge connection, while $\bD_\mu$ is the most general one, including
the gauge bundles as well. Although this may sound like a little bit of overkill,
these painstaking distinctions should prove  helpful for clarifying the
geometry of the spinors.

\subsection{Vielbein and the Kosmann Lift}

Before we get to the matter of the Lie derivative, we shall first make a lightening
review of the Cartan-Maurer formulation where the role of metric is replaced by
vielbein and the Christoffel connection by the spin connection.

For this, it is useful to introduce two separate notations for the covariant derivative.
$\nabla_\mu$
is the usual one that contains the Levi-Civita connection  {$\Gamma^\lambda_{\;\;\mu\nu}$}
and acts on the coordinate indices. When we introduce the vielbein $e^a_{\;\,\mu}$ and
its inverse $e_a^{\;\,\mu}$ such that
\bea
g_{\mu\nu} =\eta_{ab}e^a_{\;\,\mu}e^b_{\;\,\nu}\ ,\qquad g^{\mu\nu} =\eta^{ab}e_a^{\;\,\mu}e_b^{\;\,\nu} \ ,
\eea
it follows
\bea
\nabla_\mu  e^a_{\;\,\nu} = -w_{\mu\;\;b}^{\;\,a}\, e^b_{\;\nu}
\eea
for some matrix-valued $w_\mu$. The orthonormality of the vielbein
\bea
g^{\mu\nu}e^a_{\;\,\mu}e^b_{\;\,\nu} =\eta^{ab}
\eea
implies
that $w_\mu$ is $SO$-valued and thus antisymmetric with respect to its two local Lorentz indices.

On the other hand, there is a natural extension of $\nabla$ to include the action on
the local Lorentz indices, to be denoted as $D$ here, such that
\bea
D_\mu e^a_{\;\,\nu}\equiv \nabla_\mu e^a_{\;\,\nu} +w_{\mu\;\;b}^{\;\,a}\, e^b_{\;\,\nu}=0\ .
\eea
Recall that
the antisymmetrization of the two coordinate indices in the last vanishing equation
is what leads to the usual torsion-free condition,
\bea\label{torsion_free}
de^a +w^a_{\;\;b}\wedge e^b=0\ ,
\eea
which is the starting point of the Cartan-Maurer formulation of the Riemannian geometry.

Starting from (\ref{torsion_free}), one can derive the explicit expression for
$\boldsymbol\delta w$ induced by an arbitrary variation of the vielbein $\boldsymbol\delta e$
\bea
\boldsymbol\delta w_{\lambda\,bc}
&=&
-\left(
\delta\indices{^\rho_\lambda}\delta\indices{^a_{\!\![b}}e_{c]\nu}^{}
+e\indices{_{[c}^\rho}e\indices{^a_{\!\!|\lambda|}}e_{b]\nu}^{}
-e\indices{_{[b}^\rho}\delta\indices{^a_{c]}}g_{\lambda\nu}
\right)
\!{D_\rho\boldsymbol\delta e\indices{_a^\nu}}
\nonumber\\[1.5ex]
&=&-(\cC\indices{^{\rho a}_\nu})_{\lambda[bc]}^{}\,
{D_\rho\boldsymbol\delta e\indices{_a^\nu}} \ ,
\eea
with the same tensor $\cC$ defined in (\ref{cC}), some of whose coordinate indices
are swapped in favor of the local Lorentz indices. This will become useful when we derive
the symmetric energy-momentum tensor from the vielbein variation in the next section.

%\bea
%\gamma^{bc}\delta w_{\lambda\, bc}
%&=& \gamma^{\rho\sigma}(e_{a\sigma} D_\lambda \delta e\indices{^a_\rho}+e_{a\rho}D_\sigma\delta e\indices{^a_\lambda}-e_{a\lambda}D_\rho\delta e\indices{^a_\sigma})\cr\cr
%&=&\gamma^{\rho\sigma}(\cC^{\mu\nu\tau})_{\lambda\rho\sigma} e_{a\tau}D_\mu \delta e\indices{^a_\nu}
%\eea

Just as the metric can be used to raise and lower indices including those attached to
$\nabla$, the vielbein and its inverse can be used to convert between the local Lorentz indices
and the coordinate induces, provided that $D$ is used universally as the covariant derivative.
In this sense, $D$ is the correct connection to be used when we introduce the frame bundle
defined through the vielbein in addition to general tensor products of the tangent and
the co-tangent bundles. As such, we can also use $D_a$ consistently by writing
\bea
D_a \equiv e_a^{\;\,\mu} D_\mu\ , \qquad D^a \equiv e^{a\mu} D_\mu\ ,
\eea
since $D_\mu$ commutes with $e$'s. $D$ is inclusive of $\nabla$ in the sense, for example,
\bea
D_a v^b= {e_a^{\;\,\mu}}D_\mu v^b=e_a^{\;\,\mu}e^b_{\;\,\nu}\nabla_\mu v^\nu \ ,
\eea
etc.

Now the question is how we should extend the Lie derivative $\mathfrak L_\xi$
when acting on objects with the local Lorentz indices and by inference on
those with spinor indices. There appear to be two distinct choices
found in the literature. These two choices differ by a local Lorentz rotation, denoted below as $\hat\xi_K$,
itself determined from $\xi$. Since the latter is an additional gauge redundancy
that arises and partially cancels out the larger number of components of $e^a_{\;\,\mu}$
over $g_{\mu\nu}$, one may consider it a matter of choice. However, the frame bundle
built up from this local Lorentz rotation is not trivial but rather glued to
that of the tangent bundle, which suggests that the two may not be
choices on an equal footing.

One simple-minded attempt is to take the attitude that $e^a_{\;\,\mu}$ is merely a set of 1-forms
with an additional label. This would lead to
\bea
\mathfrak L_\xi e^a =d(\xi\lrcorner\,e^a)+\xi\lrcorner\, d e^a \ ,
\eea
just as on any other 1-form. If we take this attitude, one would end up with
\bea
\mathfrak L_\xi v^a=\xi^\mu\partial_\mu v^a\ ,
\eea
where $v^a\equiv  e^a_{\;\,\nu}v^\nu$ is a vector written in the vielbein basis,
for example.
This choice $\mathfrak L_\xi$ treats the coordinate indices and the
local Lorentz indices very differently. This together with its natural
extension to the spinors is in effect the choice made in many supergravity
literature. We will call it the vanilla Lie derivative in this note,
for the lack of a well-established handle.

There is an alternative to $\mathfrak L_\xi$, to be denoted as $\mathscr L_\xi $,
with better geometric motivation, called the Kosmann lift \cite{Kosmann:1971},
\bea
%\boldsymbol\delta_\xi e^a_{\;\,\mu} \!&=&\!
\mathscr{L}_\xi e^a_{\;\,\mu} &\equiv &
  {D_\mu\xi^a}  - \hat\xi^{ab}_V e_{b\mu} \;=\;
 {D^{(a} \xi^{b)} e_{b\mu}}
 \ , \cr\cr
%\boldsymbol\delta_\xi e_b^{\;\,\mu} \!&=&\!
\mathscr{L}_\xi e_b^{\;\,\mu}&\equiv&
%= - (D_b \xi^\mu) -(\hat\xi_V)_{bc}e^{c\mu}
- D_b \xi^\mu + (\hat\xi_V)_{cb}e^{c\mu} \; =\;
 {-D_{(b} \xi_{c)} e^{ c\mu}} \ ,
\eea
with
\bea
\hat\xi^{ab}_V \equiv D^{[b}\xi^{a]}\ ,
\eea
which represents an additional local Lorentz rotation.

Using $De=0$ and defining
\bea\label{V2K}
\hat\xi^{ab}_K \equiv \hat\xi^{ab}_{V}-\xi^\lambda w_{\lambda}^{\;\;ab}\ ,
\eea
the same can be rewritten as
\bea
\mathscr{L}_\xi e^a_{\;\,\mu}&=&
\xi^{\lambda} D_{\lambda} e^a_{\;\,\mu}+ D_{\mu} \xi^{\lambda} e^a_{\;\,\lambda}  - \hat\xi_{V}^{ab}e_{b\mu}\cr\cr
&=& \xi^{\lambda} \nabla_{\lambda} e^a_{\;\,\mu}+ \nabla_{\mu} \xi^{\lambda} e^a_{\;\,\lambda}  - \hat\xi_{K}^{ab}e_{b\mu}
\;=\;\mathfrak{L}_\xi e^a_{\;\,\mu}  - \hat\xi_{K}^{ab}e_{b\mu} \ ,
\eea
etc. The last expression tells us that this Kosmann lift is an extension of the
vector field $\xi$, a section of the tangent bundle, to the frame bundle as
\bea\label{Kosmann_Lift}
\xi =\xi^\mu\frac{\partial}{\partial x^\mu}\quad \rightarrow \quad
\xi^\mu\frac{\partial}{\partial x^\mu}-\hat\xi_K^{ab}\mathbf J_{ab}\ ,
\eea
where $\mathbf J_{ab}$ is the generator of the local Lorentz rotation.

This gives, in particular,
\bea\label{Kosmann_v}
\mathscr{L}_\xi v^a\;=\; \xi^b D_b v^a - \hat\xi^{ab}_{V} v_b \;=\; (\mathfrak{L}_\xi v)^\mu e^a_{\;\,\mu} + D^{(b} \xi^{a)} v_b \ ,
\eea
with the last piece distinguishing the two versions of
the Lie derivatives. The special nature of the Kosmann lift resides here,
in fact. Imagine a pair of Killing vectors $\xi$ and $\zeta$.
Even if one insists on using the vielbein as the basis, we find that
\bea
[\mathscr{L}_\xi, \mathscr{L}_\zeta ]= \mathscr{L}_{[\xi,\zeta]}\ ,
\eea
under the Kosmann lift. In this sense, the Kosmann lift is the natural
extension of the Lie derivative on (co-)tangent bundles to the frame
bundle that respects both the general covariance and the commutator
algebra among isometries.

\subsection{Kosmann Lift on Spinors}

All of these have one more step to go for  the spinor bundle as well.
The covariant derivative, for example, would be extended to
\bea
\mathscr D_\mu = D_\mu +\frac14 w_{\mu\,ab}\gamma^{ab}\ ,
\eea
such that the Dirac operator is $\gamma^a e_a^{\;\,\mu}\mathscr D_\mu$.
In particular, the requisite property
\bea
\mathscr D_\mu(\gamma^a)=
[\mathscr D_\mu,\gamma^a]=0
\eea
emerges naturally with this $\mathscr D_\mu$, with the action of $D_\mu$ on the
orthonormal indices undone by the commutator against $w_{\mu\,ab}\gamma^{ab}/4$.

The
covariant derivative inside the Dirac operator is sometimes written as
\bea
\partial_\mu +\frac14 w_{\mu\,ab}\gamma^{ab}
\eea
which, for example, appears naturally as a covariantized conjugate momentum in the supersymmetric non-linear
sigma models. When $\mathscr D_\mu$ acts directly on spinors, with no other spacetime index, the two
are identical. Although this is a matter of choice, the latter can easily become cumbersome for
more general computation, since it would not annihilate $\gamma^a$'s.

Since the spinor bundle is built upon the frame bundle, the Kosmann lift of the
vielbein implies a particular extension of the Lie derivative to spinors as well.
It is not difficult to work out the action by noting that a spinor can generate
antisymmetric tensors in the local Lorentz frame $\bar \Psi\gamma^{a\cdots b}\Psi$.
This results in,\footnote{The same was also independently discovered in the physics
side on and off with one of the earliest such due to Jackiw and Manton \cite{Jackiw:1979ub}. For later
literature with a nod to Kosmann plus more physics-friendly presentations, readers are
referred to Refs.~\cite{Figueroa:1999,Figueroa-OFarrill:2007omz}, albeit with different spinor
conventions than ours, and also to Ref.~\cite{Tomasiello}.
More recent examples where the Kosmann lift for spinors and vielbeins was employed crucially for various physics
applications may be found, for instance, in Refs.~\cite{Kelekci:2014ima,Jacobson:2015uqa,Coimbra:2016ydd}.}
\bea
 \mathscr{L}_\xi \Psi \equiv  \xi^\mu {\mathscr D}_\mu \Psi -\frac14 \hat\xi^{ab}_{V}\gamma_{ab}\Psi
=  \xi^\mu\partial_\mu \Psi -\frac14 \hat\xi^{ab}_{K}\gamma_{ab}\Psi    \ ,
\eea
which is in accord with (\ref{Kosmann_Lift}), and also leads to the transformation of
the spin connection as
\bea
\boldsymbol\delta_\xi  w = d_w \hat\xi_V+\xi\lrcorner \;\cR = d_w \hat\xi_K + \mathfrak{L}_\xi w \ ,
\eea
where $d_w$ is the covariant exterior differential
\bea
d_w \hat\xi\equiv d\,\hat\xi +w\,\hat\xi-\hat\xi \,w
\eea
acting on $\hat\xi$ that denotes either of the two matrix-valued 1-forms, $\hat\xi_{V}$ and $\hat\xi_{K}$.

A big advantage of the local Lorentz indices over the coordinate index is how
the Dirac matrices $\gamma^a$, which are constant and also $\mathscr D_\mu$-covariantly
constant, are also inert under $\mathscr L_\xi$,
\bea
\mathscr L_\xi(\gamma^a) = 0\ ,
\eea
which can be understood from how the rotation by $\hat\xi_V$ acts
twice, once by rotation the local Lorentz index and once more by
commutator action against  $-\hat\xi^{ab}_{V}\gamma_{ab}/4$. The
two cancel each other out, for the same reason as how the spin connections in $\mathscr D_\mu$
do not rotate $\gamma^a$ in the end.

\subsection{A Look-Back on the Vanilla Lie Derivative }

The Kosmann lift is the natural extension of vector fields on $\cM_d$,
a section of the tangent bundle thereof, to the frame bundle. On the other hand,
if one is merely interested in reaching the right Lie derivative
structures for tensors with coordinate indices only, the rotation of the orthonormal
indices by $-\hat\xi_K$ may appear extraneous and  even  irrelevant.
This rotation will drop out eventually, one would think, if
all such orthonormal indices and spinor indices are contracted away leaving
behind the coordinate indices only.

As such, we seem to have an option of dropping
$\hat\xi_K$ in (\ref{Kosmann_Lift}), reverting back to $\mathfrak L_\xi$,
\bea\label{vanilla_Lie}
\boldsymbol\delta'_\xi e^a \equiv\mathfrak L_\xi e^a =\xi\lrcorner\, de^a +  d(\xi^a)\ ,
\eea
and similarly,
\bea
\boldsymbol\delta'_\xi \Psi \equiv\mathfrak L_\xi\Psi =\xi^\mu\partial_\mu \Psi\ , \qquad %\mathfrak L_\xi v^a =\xi^\mu\partial_\mu v^a\ ,
\boldsymbol\delta'_\xi  w \equiv \mathfrak{L}_\xi w \ ,
\eea
where we treat the spinor indices and the orthonormal indices as
mere extra labels for these multi-component functions and  1-forms. Needless to
say,
\bea
\mathfrak L_\xi(\gamma^a)=\xi^\mu\partial_\mu(\gamma^a)=0
\eea
holds naturally. This is the choice often found in many physics contexts, especially
in the supergravity texts.

Recall that this transformation can arise as a result from a passive transformation
 $\tilde x(x)=x- \epsilon \xi(x)$ with  $\epsilon\ll 1$. The local coordinate transformation
needs to make sense only locally, chart by chart, and is thus
not intended  to extend to the entire manifold. At first sight, there
appears to be no reason not to use this vanilla Lie derivative
$\mathfrak L_\xi$ when it comes to such passive and local coordinate transformations.
On the other hand, the $SO$ structure of the orthonormal frames is eventually tied to the tangent
indices with $De^a=0$,  so that the spin connection cannot be independent
of the Levi-Civita connection.

This tells us that
although the above vanilla Lie derivative $\mathfrak L_\xi$ is available locally, it won't generally
extend to the entire manifold.
One place where we can see this most clearly is (\ref{V2K}). Turning off the Kosmann lift
means $\hat\xi^{ab}_K =0$ which translates to the condition,
\bea
\xi^\lambda w_{\lambda}^{\;\;ab}= -D^{[a}\xi^{b]}\ .
\eea
This equates the covariant expression on the right to a non-covariant one on the left, which
signals that the vanilla Lie derivative $\mathfrak L_\xi$ on spinor can be neither extended
covariantly beyond an immediate local neighborhood nor defined in a frame-independent manner.

Another fatal problem with this transformation in the physics context is how a diffeomorphism
preserves the Lagrangian $\cL$ up to a total derivative $\sim d(\xi\lrcorner\,\cV \cL)$,
so that the statement that a theory is generally covariant always relies on Stokes' theorem.
In order to argue the total derivative away, either on a compact spacetime or via appropriately
vanishing boundary condition, we should be able to extend the transformation on individual
fields everywhere on the manifold. If a transformation of a field
fails to extend beyond an immediate neighborhood, the use of Stokes' theorem becomes a little
strange, so the very statement that $\boldsymbol\delta'_\xi$ generates a symmetry of an
action principle does not make much sense.

With the  $SO$ rotation by $\hat\xi_K$, on the other hand,  the Kosmann-lifted
Lie derivative $\mathscr L_\xi$ may be glued
between overlapping local neighborhoods and would make sense globally, as long
as the vector field $\xi$ being used is defined globally.
What we will find in the next section is that this Kosmann lift also offers a clear advantage
that allows us to arrive at a sensible energy-momentum tensor, both in the Noether side
$\hat{\mathbbm{T}}$ and also in its geometric counterpart $T$ and that eventually $T=\hat\iT$
emerges naturally on par with the case of bosonic fields.

\section{Spinors with the Kosmann Lift}

Let us now turn to the question of the energy-momentum and the conservation law
for theories involving spinors, for which the vielbein and the spin connection
are indispensable. We will consider a general Lorentz-invariant  {Lagrangian} density of type
\bea
\cV\cL \;=\;d^{\,d}x \:{\rm det}(e^a_{\;\,\mu})\;\cL(\Psi,\mathscr D\Psi;e)\ ,
\eea
with $\gamma^a$'s, which are both constant and $\mathscr D_\mu$-covariantly constant,
understood for the construction of $\cL$. We shall consider its variation under $\boldsymbol\delta_\xi$ or
$\boldsymbol\delta'_\xi $. Either way, the  net variation is again a total derivative
\bea
d(\xi\lrcorner\,\cV\cL)\ .
\eea
The question comes down to how we should split these into two mutually-canceling parts,
and to what forms of these energy-momentum tensors, $\hat{\mathbbm{T}}$ and $T$,
emerge naturally depending on the choice between the two versions of the Lie derivative.

One central fact is how the Kosmann-lift of the Lie
derivative on the vielbein involves symmetric combinations in that
\bea
\boldsymbol\delta_\xi e^a_{\;\,\mu}= \mathscr L_\xi e^a_{\;\,\mu}= D^{(b} \xi^{a)} e_{b\mu}\ ,\qquad
\boldsymbol\delta_\xi e_b^{\;\,\mu}= \mathscr L_\xi e_b^{\;\,\mu}= -D_{(b} \xi_{c)} e^{c\mu}\ .
\eea
Given this, the variation of the vielbein and the spin connection
would yield
\bea
\sim (\mathscr L_\xi e_b^{\;\,\mu})(-T^b_{\;\;\mu}) =D_{(b} \xi_{c)}T^{bc}
\eea
in the end, with symmetric tensor $T^{bc}$ emerging naturally for any
$\cL(\Psi,\mathscr D\Psi;e)$.
This is entirely analogous to how $\mathfrak L_\xi g^{\mu\nu}
=-(\nabla^\mu\xi^\nu+\nabla^\nu\xi^\mu)$ enters crucially as in
\bea
\sim (\mathfrak L_\xi g^{\mu\nu}) (-T_{\mu\nu}/2)= \nabla^{(\mu}\xi^{\nu)} T_{\mu\nu}
\eea
for any Lagrangian involving scalars and tensors.

For both, the conservation law involves a symmetric energy-momentum tensor, signaling
that the closest analog of $\mathfrak L_\xi g^{\mu\nu}$
resides in the Kosmann-lifted Lie derivative on the vielbein $\mathscr L_\xi e_b^{\;\,\mu}$. Also, the
same quantity will appear as the source term for the Einstein equation.
Eventually, we will see that the Kosmann lift naturally brings us to the conservation
law of this symmetric energy-momentum tensor, with no ``improvement" involved,
regardless of whether we follow a Noether procedure or vary the vielbein.
This should be compared to the vanilla Lie derivative of the vielbein, say,
\bea
\boldsymbol\delta'_\xi e_b^{\;\,\mu} =\mathfrak L_\xi e_b^{\;\,\mu}
=\xi^\nu\partial_\nu e_b^{\;\,\mu} -(\partial_\nu\xi^\mu)e_b^{\;\,\nu}\ ,
\eea
which exhibits no obvious combinatoric property. The analog of $-T^b_{\;\;\mu}$
multiplying this version would not elevate to a symmetric tensor when the lower
coordinate index is raised to an upper local Lorentz index.

\subsection{Symmetric Energy-Momentum Tensor $T^{ab}$}

Since the covariant derivative $D$ is compatible with both the coordinate indices and the
local Lorentz indices, we will  use $D$ in place of $\nabla$ even when only
coordinate indices are present, from now on. On spinors, on the other hand,
we will insist on the notation $\mathscr D$ to emphasize the spinor indices yet to be contracted away.
Needless to say, these covariant derivatives are defined with respect to the unperturbed vielbein.

Let us consider a Dirac field $\Psi$ and the action in the form
\bea
\cL(\Psi,\sD_\lambda\Psi,\bar\Psi,\sD_\lambda\bar\Psi;e\indices{_a^\nu})\ ,
\eea
where $\sD_\lambda\bar\Psi$ is a shorthand notation for $(\sD_\lambda\Psi)^\dagger\gamma^0$
in the Lorentzian signature.
The symmetric energy-momentum follows from the variation of the matter action with respect to $e\indices{_a^\nu}$, i.e.,
\bea
\frac{\boldsymbol\delta(\cV\cL)}{\boldsymbol\delta e\indices{_a^\nu}}\boldsymbol\delta e\indices{_a^\nu}
\;=\;\cV
\left(
-e\indices{^a_\nu}\cL
+\frac{\boldsymbol\delta\cL}{\boldsymbol\delta e\indices{_a^\nu}}
\right)
\boldsymbol\delta e\indices{_a^\nu} \ ,
\eea
and the second term includes pieces that come from the variation of the connection,
\bea
%\frac{\boldsymbol\delta\cL}{\boldsymbol\delta e\indices{_a^\nu}}
%\boldsymbol\delta e\indices{_a^\nu}
%\!\!&=&\!\!
%\frac{\boldsymbol\partial\cL}{\boldsymbol\partial e\indices{_a^\nu}}
%\boldsymbol\delta e\indices{_a^\nu}
%+\frac{\boldsymbol\partial\cL}{\boldsymbol\partial(\sD_\lambda\Psi)}
%(\boldsymbol\delta\sD)_\lambda\Psi
%+(\boldsymbol\delta\sD)_\lambda\bar\Psi
%\frac{\boldsymbol\partial\cL}{\boldsymbol\partial(\sD_\lambda\bar\Psi)}
%\cr\cr
%\!\!&=&\!\!
%\frac{\boldsymbol\partial\cL}{\boldsymbol\partial e\indices{_a^\nu}}
%\boldsymbol\delta e\indices{_a^\nu}+
\frac{1}{4}\boldsymbol\delta w_{\lambda\,bc}
\left(
\frac{\cev{\boldsymbol\partial}\cL}{\boldsymbol\partial(\sD_\lambda\Psi)}\gamma^{bc}\Psi
-\bar\Psi\gamma^{bc}\frac{\vec{\boldsymbol\partial}\cL}{\boldsymbol\partial(\sD_\lambda\bar\Psi)}
\right) \ , \quad
\boldsymbol\delta w_{\lambda\,bc}
%&=&
%-\left(
%\delta\indices{^\rho_\lambda}\delta\indices{^a_{\!\![b}}e_{c]\nu}^{}
%+e\indices{_{[c}^\rho}e\indices{^a_{\!\!|\lambda|}}e_{b]\nu}^{}
%-e\indices{_{[b}^\rho}\delta\indices{^a_{c]}}g_{\lambda\nu}
%\right)
%\!{D_\rho\boldsymbol\delta e\indices{_a^\nu}}
%\nonumber\\[1.5ex]
=-(\cC\indices{^{\rho a}_\nu})_{\lambda[bc]}^{}\,
{D_\rho\boldsymbol\delta e\indices{_a^\nu}} \ ,
\eea
with the same tensor $\cC$ defined in (\ref{cC}), now written with both the local Lorentz indices, $a,b,\cdots$ and the coordinate
indices, $\rho, \nu,\cdots$.

After much  manipulation, we end up with
\begin{align}
&\frac{\boldsymbol\delta(\cV\cL)}{\boldsymbol\delta e\indices{_a^\nu}}
\boldsymbol\delta e\indices{_a^\nu}
\nonumber\\[0.5ex]
&\!\!=\;\cV
\left[
-\Sigma\indices{^a_\nu}\boldsymbol\delta e\indices{_a^\nu}
-\frac{1}{4} D_\rho\!
\left[
(\cC\indices{^{\rho a}_\nu})_{\lambda[bc]}^{}\boldsymbol\delta e\indices{_a^\nu}\!
\left(	
\frac{\cev{\boldsymbol\partial}\cL}{\boldsymbol\partial(\sD_\lambda\Psi)}\gamma^{bc}\Psi
-\bar\Psi\gamma^{bc}\frac{\vec{\boldsymbol\partial}\cL}{\boldsymbol\partial(\sD_\lambda\bar\Psi)}
\right)
\right]
\right] ,
\end{align}
where
\bea\label{EMSigma}
\Sigma\indices{^a_\nu}
\;\equiv\;e\indices{^a_\nu}\cL
-\frac{\boldsymbol\partial\cL}{\boldsymbol\partial e\indices{_a^\nu}}
-\frac{1}{4}(\cC\indices{^{\rho a}_\nu})_{\lambda[bc]}^{} D_\rho\!
\left(
\frac{\cev{\boldsymbol\partial}\cL}{\boldsymbol\partial(\sD_\lambda\Psi)}\gamma^{bc}\Psi
-\bar\Psi\gamma^{bc}\frac{\vec{\boldsymbol\partial}\cL}{\boldsymbol\partial(\sD_\lambda\bar\Psi)}
\right) \ .
\eea
Let us split the tensor $\Sigma$ into
\bea
\Sigma^{ab}= T^{ab}+\Sigma^{[ab]}\ ,
\eea
where the symmetric part is now called $ T^{ab}$, suggestively.

Now, let us finally specialize the variation to that of the infinitesimal diffeomorphism with the
Kosmann lift, $\boldsymbol\delta e\indices{_a^\nu}\rightarrow \boldsymbol\delta_\xi e\indices{_a^\nu}=\mathscr L_\xi e\indices{_a^\nu}=-D_{(a}\xi_{b)}e^{b\nu}$.
The variation reduces to
\bea\label{Symmetric_Dirac}
\frac{\boldsymbol\delta(\cV\cL)}{\boldsymbol\delta e\indices{_a^\nu}}\boldsymbol\delta_\xi e\indices{_a^\nu}
=\cV\left[(D_a\xi_b)T^{ab}+D_\rho\,\cU^\rho\right] \ ,
\eea
with the symmetric parts only,  which implies the conservation law
\bea
D_aT^{ab}=0\ ,
\eea
upon integrating by parts and throwing away the two total derivative terms. This
motivates us to identify $T^{ab}$ as the energy-momentum tensor. The analog of $\cS$
found in (\ref{Symmetric_Mixed}) is
\bea\label{Symmetric_Dirac_U}
\cU^\rho
=-\frac{1}{2} D_{(\mu}\xi_{\nu)}
\left(
\frac{\cev{\boldsymbol\partial}\cL}{\boldsymbol\partial(\sD_\nu\Psi)}\gamma^{\rho\mu}\Psi
-\bar\Psi\gamma^{\rho\mu}\frac{\vec{\boldsymbol\partial}\cL}{\boldsymbol\partial(\sD_\nu\bar\Psi)}
\right) \ .
\eea

\subsubsection*{What does $\Sigma^{[ab]}$ do?}

A puzzling left-over from the above is the antisymmetric part $\Sigma^{[ab]}$. It enters via the
general variation of the vielbein, but drops out by the time we specialize to the
diffeomorphism variation, $\boldsymbol\delta_\xi e\indices{_a^\nu}$, provided that we use the Kosmann
lift for $\boldsymbol\delta_\xi  e\indices{_a^\nu} =\mathscr L_\xi e\indices{_a^\nu}$.
There is no conservation law on this quantity unlike the symmetric part $T^{ab}$.

On the other hand, the general variation with respect to the vielbein
in the presence of the Einstein-Hilbert action produces the Einstein equation,
\bea
G_{ab}=8\pi G_N T_{ab}\ .
\eea
The symmetric property of the Einstein tensor on the left again projects $\Sigma^{[ab]}$ out
from the Einstein equation, but this seemingly implies extra equation,
\bea\label{vanishing_Sigma}
0=\Sigma^{[ab]}\ ,
\eea
which would be strange. Thankfully, the structure of $\Sigma^{[ab]}$ has a very special form.

Recall how the local Lorentz symmetry that rotates the orthonormal indices
is much like an internal gauge symmetry, needed to match the
different numbers of components between the metric and the vielbein. This
is represented by a position-dependent $SO$ matrix, say $e^{L}$, acting
on orthonormal indices and by inference on spinor indices. With arbitrary anti-symmetric
matrix $L_{ab}$, we find immediately,
\bea
L_{ab}\Sigma^{[ab]}=
\left(
\frac{\cev{\boldsymbol\partial}\cL}{\boldsymbol\partial\Psi}
-\sD_\lambda\frac{\cev{\boldsymbol\partial}\cL}{\boldsymbol\partial(\sD_\lambda\Psi)}
\right)
\boldsymbol\delta_L\Psi
+\boldsymbol\delta_L\bar\Psi
\left(
\frac{\vec{\boldsymbol\partial}\cL}{\boldsymbol\partial\bar\Psi}
-\sD_\lambda\frac{\vec{\boldsymbol\partial}\cL}{\boldsymbol\partial(\sD_\lambda\bar\Psi)}
\right)\ ,
\eea
from the point-wise invariance of $\cL$ under the local Lorentz transformation.
%\bea
%\Sigma^{[ab]}=
%\left(
%\frac{\cev{\boldsymbol\partial}\cL}{\boldsymbol\partial\Psi}
%-\sD_\lambda\frac{\cev{\boldsymbol\partial}\cL}{\boldsymbol\partial(\sD_\lambda\Psi)}
%\right)
%\!\frac{\gamma^{ab}}{4}\Psi
%-\bar\Psi\frac{\gamma^{ab}}{4}\!
%\left(
%\frac{\vec{\boldsymbol\partial}\cL}{\boldsymbol\partial\bar\Psi}
%-\sD_\lambda\frac{\vec{\boldsymbol\partial}\cL}{\boldsymbol\partial(\sD_\lambda\bar\Psi)}
%\right)
%\eea
This immediately signals that this quantity vanishes on-shell classically, so
that (\ref{vanishing_Sigma}) does not incur a new equation of motion.

However, this may not suffice since we could imagine quantizing matter fields in
the presence of the classical gravity, whereby we should find
\bea
G_{ab}=8\pi G_N\boldsymbol\langle T_{ab}\boldsymbol\rangle\ ,
\eea
which leaves behind a condition of the vanishing expectation value $\boldsymbol\langle \Sigma^{[ab]}\boldsymbol\rangle =0$.
Is this automatic at quantum level, as with the classical counterpart? What comes to the rescue is again the local Lorentz symmetry.

With dynamical gravity present, classical or quantum, a natural expectation is
that the diffeomorphism anomalies cancel out one way or another. At quantum level,
this is an obvious requirement since otherwise the quantization itself is in danger.
Less noticed is that the same is needed for the above semiclassical Einstein equation
as well, since the Einstein tensor is divergence-free as a mathematical identity
which makes sense only if the right hand side is also divergence-free
In turn this translates to how the local Lorentz symmetry is anomaly-free.

The measure of the fermions in the path integral must be therefore by itself invariant
under the local Lorentz rotation, or, if not, the anomalous phase should be canceled
by some anomaly inflow. Recall how the dynamical fields
are dummy variables to be integrated over,  so that
\bea
\int[D\Psi \cdots]\,e^{\ii S(\Psi,\cdots) }=\int[D(\Psi+\epsilon\boldsymbol\delta\Psi) \cdots]\,e^{\ii S(\Psi+\epsilon\boldsymbol\delta\Psi,\cdots) }
\eea
for any $\boldsymbol\delta$. The invariance of the $\Psi$-measure  by itself, or
possibly together with an inflow contribution, will enforce
\bea
0=\boldsymbol\langle L_{ab} \Sigma^{[ab]}\boldsymbol\rangle\ ,
\eea
bringing us back to the desired vanishing of $\boldsymbol\langle \Sigma^{[ab]}\boldsymbol\rangle$.

\subsection{Noether Energy-Momentum $\hat{\mathbbm{T}}\indices{^\mu_\alpha}$}

For a Dirac field $\Psi$, the Noether procedure is to be performed using
\bea
\boldsymbol\delta_\xi\Psi
&=&\xi^\alpha\sD_\alpha\Psi
+\frac{1}{4}( D_\rho\xi_\sigma)\gamma^{\rho\sigma}\Psi\ ,
\cr\cr
\boldsymbol\delta_\xi\bar\Psi
&=&\xi^\alpha\sD_\alpha\bar\Psi
-\frac{1}{4}( D_\rho\xi_\sigma)\bar\Psi\gamma^{\rho\sigma} \ ,
\eea
again with all gauge fields coupled to $\Psi$ making their natural appearances in $\mathscr D$ as well,
as we noted at the head of this section.

With the Lagrangian in the form $\cL(\Psi,\sD\Psi,\bar\Psi,\sD\bar\Psi;e)$, we have
\bea
\!\!\!\!\!
\!\!\!\!\!
\boldsymbol\delta_\xi\cL\,\biggr\vert_{e\;{\rm fixed}}
\!&=&\!\frac{\cev{\boldsymbol\partial}\cL}{\boldsymbol\partial\Psi}\boldsymbol\delta_\xi\Psi
+\frac{\cev{\boldsymbol\partial}\cL}{\boldsymbol\partial(\sD_\lambda\Psi)}\sD_\lambda(\boldsymbol\delta_\xi\Psi)
+\boldsymbol\delta_\xi\bar\Psi\frac{\vec{\boldsymbol\partial}\cL}{\boldsymbol\partial\bar\Psi}
+\sD_\lambda(\boldsymbol\delta_\xi\bar\Psi)\frac{\vec{\boldsymbol\partial}\cL}{\boldsymbol\partial(\sD_\lambda\bar\Psi)}
\cr\cr
\!&=&\!\xi^\alpha\sD_\alpha\cL
+\frac{\cev{\boldsymbol\partial}\cL}{\boldsymbol\partial(\sD_\lambda\Psi)}
\left(
\xi^\alpha[\sD_\lambda,\sD_\alpha]\Psi
+\frac{1}{4}( D_\lambda D_\rho \xi_\sigma)\gamma^{\rho\sigma}\Psi
\right)
\cr\cr
\!&&\!+
\left(
\xi^\alpha[\sD_\lambda,\sD_\alpha]\bar\Psi
-\frac{1}{4}( D_\lambda D_\rho \xi_\sigma)\bar\Psi\gamma^{\rho\sigma}
\right)
\frac{\vec{\boldsymbol\partial}\cL}{\boldsymbol\partial(\sD_\lambda\bar\Psi)}
+ (D_\mu\xi^\alpha) \cK^\mu_{\;\;\alpha} \ ,
\eea
with
\bea
\cK^\mu_{\;\;\alpha}&\equiv&
\frac{1}{4}\!
\left(
\frac{\cev{\boldsymbol\partial}\cL}{\boldsymbol\partial\Psi}\gamma\indices{^\mu _\alpha}\Psi
-\bar\Psi\gamma\indices{^\mu _\alpha}\frac{\vec{\boldsymbol\partial}\cL}{\boldsymbol\partial\bar\Psi}
+\frac{\cev{\boldsymbol\partial}\cL}{\boldsymbol\partial(\sD_\lambda\Psi)}\gamma\indices{^\mu_\alpha}\sD_\lambda\Psi
-\sD_\lambda\bar\Psi\gamma\indices{^\mu_\alpha}\frac{\vec{\boldsymbol\partial}\cL}{\boldsymbol\partial(\sD_\lambda\bar\Psi)}
\right)
\cr\cr
&&+\frac{\cev{\boldsymbol\partial}\cL}{\boldsymbol\partial(\sD_\mu\Psi)}\sD_\alpha\Psi
+\sD_\alpha\bar\Psi\frac{\vec{\boldsymbol\partial}\cL}{\boldsymbol\partial(\sD_\mu\bar\Psi)}
\ .
\eea

The commutator $[\sD,\sD]$ acting on $\Psi, \bar\Psi$ can be moved to $\xi$ as
\bea
\xi^\alpha[\sD_\lambda,\sD_\alpha]\Psi
\!&=&\!\xi^\alpha\!
\left(
\frac{1}{4}R_{\rho\sigma\lambda\alpha}\gamma^{\rho\sigma}\Psi
\right)
\;=\;\frac{1}{4}([ D_\rho, D_\sigma]\xi_\lambda)\gamma^{\rho\sigma}\Psi
\cr\cr
\xi^\alpha[\sD_\lambda,\sD_\alpha]\bar\Psi
\!&=&\!\xi^\alpha\!
\left(
-\frac{1}{4}R_{\rho\sigma\lambda\alpha}\bar\Psi\gamma^{\rho\sigma}
\right)
\;=\;-\frac{1}{4}([ D_\rho, D_\sigma]\xi_\lambda)\bar\Psi\gamma^{\rho\sigma} \ .
\eea
The term involving second derivatives of $\xi$ can be reduced as
\bea
2 D_{[\rho} D_{\sigma]}\xi_\lambda+ D_{\lambda} D_{[\rho}\xi_{\sigma]}
%&=& D_\kappa D_\mu\xi_\nu
%\left(
%2\delta\indices{^\kappa_{\!\![\rho}}\delta\indices{^\mu_{\sigma]}}\delta\indices{^\nu_\lambda}
%+\delta\indices{^\kappa_\lambda}\delta\indices{^\mu_{\!\![\rho}}\delta\indices{^\nu_{\sigma]}}
%\right)
%\cr\cr
&=& D_\kappa D_\mu\xi_\nu
\cdot2\delta\indices{^\kappa_{\!\![\rho}}\delta\indices{^{(\mu}_{\sigma]}}\delta\indices{^{\nu)}_{\!\!\lambda}} \ ,
\eea
by virtue of the symmetry of the Riemann tensor, i.e.,
\bea
 D_\kappa D_\mu\xi_\nu
\left(
2\delta\indices{^\kappa_{\!\![\rho}}\delta\indices{^{[\mu}_{\sigma]}}\delta\indices{^{\nu]}_{\!\!\lambda}}
+\delta\indices{^\kappa_\lambda}\delta\indices{^\mu_{\!\![\rho}}\delta\indices{^\nu_{\sigma]}}
\right)
%&=&\frac{1}{2}
%\left(
%[ D_{\rho}, D_{\sigma}]\xi_\lambda
%+[ D_{\sigma}, D_{\lambda}]\xi_\rho
%+[ D_{\lambda}, D_{\rho}]\xi_\sigma
%\right)
%\cr\cr
=\frac{1}{2}
\left(
R_{\rho\sigma\lambda\alpha}+R_{\sigma\lambda\rho\alpha}+R_{\lambda\rho\sigma\alpha}
\right)
\xi^\alpha=0\ .
\eea

All of these lead to
\bea
\!\!\!\!\!\!
\!\!\!\!\!\!
\boldsymbol\delta_\xi\cL\,\biggr\vert_{e\;{\rm fixed}}
\!&=&\!\xi^\alpha\sD_\alpha\cL
+(D_\mu\xi^\alpha) \cK^\mu_{\;\;\alpha}
\cr\cr
\!&&\!+\frac{1}{4}(2 D_{[\rho} D_{\sigma]}\xi_\lambda+ D_\lambda  D_{[\rho}\xi_{\sigma]})\!
\left(
\frac{\cev{\boldsymbol\partial}\cL}{\boldsymbol\partial(\sD_\lambda\Psi)}\gamma^{\rho\sigma}\Psi
-\bar\Psi\gamma^{\rho\sigma}\frac{\vec{\boldsymbol\partial}\cL}{\boldsymbol\partial(\sD_\lambda\bar\Psi)}
\right) ,
\eea
and eventually, we end up with
\bea\label{Noether_Dirac}
\frac{\cev{\boldsymbol\delta}(\cV\cL)}{\boldsymbol\delta\Psi}\boldsymbol\delta_\xi\Psi
+\boldsymbol\delta_\xi\bar\Psi\frac{\vec{\boldsymbol\delta}(\cV\cL)}{\boldsymbol\delta\bar\Psi}
\;=\;d(\xi\lrcorner\cV\cL)
+\cV\left[( D_\mu\xi^\alpha)(-\hat\iT\indices{^\mu_\alpha})
- D_\rho \,\cU^\rho \right] \ ,
%\cr\cr
%&&+\frac{1}{2}\sD_\rho\!
%\left[
% D_{(\mu}\xi_{\nu)}
%\left(
%\frac{\cev{\boldsymbol\partial}\cL}{\boldsymbol\partial(\sD_\nu\Psi)}\gamma^{\rho\mu}\Psi
%-\bar\Psi\gamma^{\rho\mu}\frac{\vec{\boldsymbol\partial}\cL}{\boldsymbol\partial(\sD_\nu\bar\Psi)}
%\right)
%\right]
%\ ,
\eea
where the Noether energy-momentum is explicitly given as
\bea\label{Noether_Spinor}
\hat\iT\indices{^\mu_\alpha}
\;=\;\delta\indices{^\mu_\alpha}\cL - \cK^\mu_{\;\;\alpha}
+\frac{1}{2}\delta\indices{^\mu_{\!\!(\lambda}}g_{\sigma)\alpha}^{} D_\rho\!
\left(
\frac{\cev{\boldsymbol\partial}\cL}{\boldsymbol\partial(\sD_\lambda\Psi)}\gamma^{\rho\sigma}\Psi
-\bar\Psi\gamma^{\rho\sigma}\frac{\vec{\boldsymbol\partial}\cL}{\boldsymbol\partial(\sD_\lambda\bar\Psi)}
\right) \ ,
\eea
and $\cU^\rho$ is the same as in (\ref{Symmetric_Dirac}).
\subsection{Equality of $\hat{\mathbbm{T}}$ and $T$}

Combining (\ref{Symmetric_Dirac}) and (\ref{Noether_Dirac}), we have
\bea
\boldsymbol\delta_\xi(\cV\cL)
&=&\frac{\boldsymbol\delta(\cV\cL)}{\boldsymbol\delta e\indices{_a^\nu}}\boldsymbol\delta_\xi e\indices{_a^\nu}
+\frac{\cev{\boldsymbol\delta}(\cV\cL)}{\boldsymbol\delta\Psi}\boldsymbol\delta_\xi\Psi
+\boldsymbol\delta_\xi\bar\Psi\frac{\vec{\boldsymbol\delta}(\cV\cL)}{\boldsymbol\delta\bar\Psi}
\cr\cr
&=&d(\xi\lrcorner\cV\cL)
+\cV\,(D_a\xi_b)(T^{ab}-\hat\iT^{ab}) +\cV\left[ \cancel{D_\rho\,\cU^\rho}-\cancel{D_\rho\,\cU^\rho}\right] \ ,
\eea
which, again with the universal property of the covariant Lagrangian density,  $\boldsymbol\delta_\xi(\cV\cL)=  d(\xi\lrcorner\cV\cL)$,
gives the desired identity,
\bea
T^{ab}=\hat\iT^{ab}\ ,
\eea
even though this may not be obvious from the respective expressions.
Again we neither relied on the equation of motion nor threw away total derivative terms along the way.

\subsection{Free Dirac Field}

The Lagrangian for the free Dirac field is given by
\bea\label{Dirac_Standard}
\cL_{\rm Dirac} \;=\; -\ii\bar\Psi\gamma^a e\indices{_a^\mu} \sD_\mu\Psi
-\ii m\bar\Psi\Psi \ .
\eea
We may use the democratic form of Lagrangian for the free Dirac field
\bea
\cL_{\rm Dirac}'&=&
\cL_{\rm Dirac}+\frac{\ii}{2}D_\mu(\bar\Psi\gamma^\mu\Psi)
\cr\cr
&=&
-\frac{\ii}{2}\!
\left(
\bar\Psi\gamma^\mu\sD_\mu\Psi
-\sD_\mu\bar\Psi\gamma^\mu\Psi
\right)
-\ii m\bar\Psi\Psi \ ,
\eea
but it turns out that either way we end up with the same energy-momentum tensor $T=\hat\iT$.

Starting from (\ref{Dirac_Standard}), we find
\bea
\cK^{\mu\nu}
&=& \frac{\ii}{4}\bar\Psi[\gamma^{\mu\nu},\gamma^\lambda]\sD_\lambda\Psi
- \ii\bar\Psi\gamma^\mu\sD^\nu\Psi
\;=\; \ii\bar\Psi\gamma^{[\mu}g^{\nu]\lambda}\sD_\lambda\Psi
- \ii\bar\Psi\gamma^\mu\sD^\nu\Psi
\cr\cr
&=& -\ii\bar\Psi\gamma^{(\mu}\sD^{\nu)}\Psi\ ,
\eea
while
\bea
&&\frac{1}{2}\delta\indices{^\mu_{\!\!(\lambda}}\delta\indices{^\nu_{\sigma)}}D_\rho
\left(
\frac{\cev{\boldsymbol\partial}\cL_{\rm Dirac}}{\boldsymbol\partial(\sD_\lambda\Psi)}\gamma^{\rho\sigma}\Psi
- \bar\Psi\gamma^{\rho\sigma}\frac{\vec{\boldsymbol\partial}\cL_{\rm Dirac}}{\boldsymbol\partial(\sD_\lambda\Psi)}
\right)\cr\cr
&=& -\frac{\ii}{4}D_\rho[\bar\Psi(\gamma^\mu\gamma^{\rho\nu}+\gamma^\nu\gamma^{\rho\mu})\Psi]
\cr\cr
&=& -\frac{\ii}{2}(\bar\Psi\gamma^{(\mu}\sD^{\nu)}\Psi + \sD^{(\nu}\bar\Psi\gamma^{\mu)}\Psi)
+ \frac{\ii}{2}g^{\mu\nu}(\bar\Psi\gamma^\rho\sD_\rho\Psi + \sD_\rho\bar\Psi\gamma^\rho\Psi)\ .
\eea
Although the three pieces of $\hat\iT$ in (\ref{Noether_Spinor}) are not in the democratic form individually,
their sum for $\hat\iT$
proves to be,
\bea\label{EM_Dirac}
\hat\iT^{ab}\!&=&\!
\eta^{ab}\!\left(
-\frac{\ii}{2}(\bar\Psi\gamma^\mu\sD_\mu\Psi-\sD_\mu\bar\Psi\gamma^\mu\Psi)-\ii m\bar\Psi\Psi
\right)
\cr\cr
\!&&\!+\frac{\ii}{2}\!
\left(\bar\Psi\gamma^{(a}\sD^{b)}\Psi-\sD^{(b}\bar\Psi\gamma^{a)}\Psi\right) \ ,
\eea
where we came back to the orthonormal indices.

Something similar happens with $T^{ab}=\Sigma^{(ab)}$ from $\cL_{\rm Dirac}$.
Three contributing pieces are all non-democratic but the last piece in (\ref{EMSigma})
reduces, upon the symmetrization of the two surviving indices,
\bea
&&-\frac{1}{4}(\cC^{\rho(ab)})_{\lambda[cd]}D_\rho
\left(
\frac{\cev{\boldsymbol\partial}\cL_{\rm Dirac}}{\boldsymbol\partial(\sD_\lambda\Psi)}\gamma^{cd}\Psi
-\bar\Psi\gamma^{cd}\frac{\vec{\boldsymbol\partial}\cL_{\rm Dirac}}{\boldsymbol\partial(\sD_\lambda\bar\Psi)}
\right)
\cr\cr
%&=& -\frac{\ii}{2}e\indices{_c^\rho}\delta\indices{^{(a}_d}e\indices{^{b)}_{\!\!\lambda}}
%D_\rho\left(\bar\Psi\gamma^\lambda\gamma^{cd}\Psi \right) \cr\cr
%&=& -\frac{\ii}{4}D_c[\bar\Psi(\gamma^a\gamma^{cb}+\gamma^b\gamma^{ca})\Psi]\cr\cr
%&=& -\frac{\ii}{2}D_c[\bar\Psi(\gamma^{(a}\eta^{b)c}-\eta^{ab}\gamma^c)\Psi]\cr\cr
&=& -\frac{\ii}{2}(\bar\Psi\gamma^{(a}\sD^{b)}\Psi + \sD^{(b}\bar\Psi\gamma^{a)}\Psi)+ \frac{\ii}{2}\eta^{ab}(\bar\Psi\gamma^c\sD_c\Psi + \sD_c\bar\Psi\gamma^c\Psi)\ ,
\eea
which induces the same expression as $T$ above. In the end, we find an expression for the
latter obeying
\bea
T^{ab}=\hat\iT^{ab}
\eea
identically, as was earlier claimed on the general ground.

If we started with the democratic $\cL_{\rm Dirac}'$, the first two contributing pieces for $\hat\iT'$ and $T'$ would be democratic to begin with, respectively, while the last pieces involving the
$\cC$-tensor would vanish identically for both $T'$ and $\hat\iT'$. These differences in the middle
steps remarkably lead to
\bea
(T')^{ab}=T^{ab}=\hat\iT^{ab}=(\hat\iT')^{ab}\ ,
\eea
despite $\cL_{\rm Dirac}'\neq \cL_{\rm Dirac}$.
This comes about in part thanks to how the content of the total derivatives, $D_\rho(\cU')^\rho$ and
$D_\rho\,\cU^\rho$, also shift between the two choices.

Note how, here, the equality $T=\hat\iT$ is seen by separate and explicit computations of the
respective formulae as well.
As noted at the head of the section, this means that the invariance of the Dirac action holds under the
combined action of the Kosmann lift $\mathscr L_\xi$ on the vielbein and the generalized Kosmann transformation
$\boldsymbol\delta_\xi$ on the spinor, confirming that they form a symmetry of the action that make sense globally.

\subsection{Generalized Kosmann with Gauge Fields}

Before we proceed further, there is one additional ingredient we need to mull over.
Spinors in field theory are often in some representations of gauge groups,
carrying additional internal indices. Under this vanilla Lie derivative, the addition of the gauge
field does not change the action on spinors; we merely need to remember that the gauge fields $\cA$
should transform by $\mathfrak L_\xi$ as well, and at least locally this suffices to guarantee the
general covariance of the Dirac action, for example. The question is if and how this situation
changes  once we adopt the Kosmann lift. What we mean by the general covariance of the
matter action is itself at stake.

Let us set our notation for the gauge sector first. From now on, we will employ the anti-hermitian
notation,
\bea
\cA=-\ii A \ , \qquad \Theta=-\ii \theta
\eea
so that
\bea\label{gauge_convention}
\boldsymbol\delta^\text{gauge}_\Theta \Psi = -\Theta \Psi \ , \qquad
\boldsymbol\delta^\text{gauge}_\Theta \cA = d\Theta +[\cA,\Theta]
\eea
and the covariant derivative,
\bea
\bD_\mu=\sD_\mu+\cA_\mu=D_\mu +\cA_\mu + \frac14 w_{\mu\,ab}\gamma^{ab} \ .
\eea
Under the Kosmann-lifted diffeomorphism,
\bea
{\boldsymbol\delta}_\xi\Psi& =& \xi^\mu\sD_\mu \Psi-\frac14 \hat\xi^{ab}_{V}\gamma_{ab}\Psi\;=\;\mathscr L_\xi \Psi\ , \cr\cr
{\boldsymbol\delta}_\xi\cA_\mu &=&\xi^\nu\partial_\nu \cA_\mu +(\partial_\mu \xi^\nu) \cA_\nu\;=\;\mathfrak L_\xi \cA_\mu
\eea
the free Dirac action from
\bea\label{Dirac_Gauged}
\cL_{\rm Dirac} \;=\; -\ii\bar\Psi\gamma^a e\indices{_a^\mu} \bD_\mu\Psi
-\ii m\bar\Psi\Psi \ ,
\eea
now equipped with the gauge-covariant derivative, is invariant in the same sense that (\ref{Dirac_Standard})
is invariant under the vanilla Lie derivative.

Being a section of the relevant vector bundle as well as a section of the spinor bundle,
on the other hand, the question of how we glue the local sections for $\Psi$ across overlapping patches
with regard to the gauge indices enters the Lie derivative also. An alternative transformation rule
for the diffeomorphism, augmented by gauge transformation by $\Theta=- (\xi\,\lrcorner\,\cA)$,
\bea
\hat{\boldsymbol\delta}_\xi\Psi &\equiv&
\left(\mathscr L_\xi + \boldsymbol\delta^\text{gauge}_{-(\xi\,\lrcorner\,\cA)}\right)\Psi
\;=\; \xi^\mu\bD_\mu \Psi -\frac14 \hat\xi^{ab}_{V}\gamma_{ab}\Psi \ ,
\cr\cr
\hat{\boldsymbol\delta}_\xi\cA_\mu
&\equiv&\left(\mathfrak L_\xi + \boldsymbol\delta^\text{gauge}_{-(\xi\,\lrcorner\,\cA)}\right)\cA_\mu
\;=\; \xi^\nu\mathcal F_{\nu\mu}
\eea
accommodates the latter need on equal footing with the spin indices.\footnote{Note that the modified transformations are now fully covariant. Such covariant combination for
$\cA_\mu$ in the conventional setting without the Kosmann lift was previously identified in Ref.~\cite{Jackiw:1978}.}

Since the difference between ${\boldsymbol\delta}_\xi$ and $\hat{\boldsymbol\delta}_\xi$
is a gauge transformation, local in the sense of a given coordinate patch,
the local covariance of the Dirac action holds equally. In addition, the latter
action makes sense globally as well, which leads to
\bea\label{generalized_Kosmann_Dirac}
\hat{\boldsymbol\delta}_\xi \left(\cV\cL_{\rm Dirac}\right)
=d\left(\xi\lrcorner\,\cV\cL_{\rm Dirac}\right)\ .
\eea
Henceforth, we will refer to the latter transformation rule $\hat{\boldsymbol\delta}_\xi$
on spinors and gauge fields as the generalized Kosmann lift.

The procedure of the previous subsections that led to the energy-momentum tensor
becomes slightly more involved, but one arrives at a straightforward generalization
of the end result~(\ref{EM_Dirac}),
\bea
T^{ab}\;=\;\hat\iT^{ab}\!&=&\!
\eta^{ab}\!\left(
-\frac{\ii}{2}(\bar\Psi\gamma^\mu\bD_\mu\Psi-\bD_\mu\bar\Psi\gamma^\mu\Psi)-\ii m\bar\Psi\Psi
\right)
\cr\cr
\!&&\!+\frac{\ii}{2}\!
\left(\bar\Psi\gamma^{(a}\bD^{b)}\Psi-\bD^{(b}\bar\Psi\gamma^{a)}\Psi\right) \ ,
\eea
with the covariant derivative upgraded from the purely gravitational $\sD$ to the gauged
$\bD$. As noted several times, the equality $T^{ab}=\hat\iT^{ab}$, computed by independent
computations on par with the previous subsections, comes about from the general covariance of the action~(\ref{generalized_Kosmann_Dirac})
and vice versa.

\section{Diffeomorphism Anomalies}

Now that we have understood how the Kosmann lift of the diffeomorphism is
essential when it comes to the energy-momentum tensor of spinors, a natural
follow-up question is what  we should do about the Ward identity. At the
most naive level, the latter
asserts that the divergence of the energy-momentum tensor must have a
vanishing expectation value. With chiral field content, however, the Ward identity can
easily fail and the quantity that replaces the zero on the other side is
called the consistent diffeomorphism anomaly.

The derivation of such anomalies is rather involved but, needless to say,
the precise transformation rules of various fields would enter centrally.
The primary examples for which the anomaly arises are Weyl fermions,
yet, we hardly hear of the Kosmann lift mentioned in
related physics literature. We wish to clear up this odd situation
by going back some forty years and retracing the steps more carefully.

In the end, we will find that the venerable anomaly polynomials are
safely reproduced, despite the generalized Kosmann lift and, in a sense to be clarified below,
thanks to the generalized Kosmann lift. However, the dictionary that extracts the
covariant diffeomorphism anomalies from the anomaly polynomial turns
out to be faulty and an additional factor 1/2 is needed. This also
affects the consistent anomaly similarly, since the usual
Wess-Zumino consistency and the subsequent anomaly descent are
homogeneous processes whose anomaly polynomials and the overall multiplicative
constant can only be fixed by the covariant side. It is likely that
in most literature this factor 1/2 is ignored or even swept under the rug by the
time the discussion reached the consistent side.

Recall how the diffeomorphism anomaly of spinors with chiral couplings is computed in two steps.
First, we compute the quantity known as the covariant anomaly. Let us denote this
quantity as $G_{\rm diff}^{\rm cov}$. One way to deal with the
latter is to mimic Fujikawa's path integral viewpoint \cite{Fujikawa:1979} and compute
\bea\label{Kosmann_Cov}
G_{\rm diff}^{\rm cov}={\rm Tr}(\Gamma \hat{\boldsymbol\delta}_\xi)
\eea
as a regulated functional trace.
One of the main lessons we learned in the previous section
is that $\boldsymbol\delta_\xi$ should be the generalized version of Kosmann-lifted
diffeomorphism. Our convention for the chirality
operator $\Gamma$ is
\bea
\Gamma \equiv %\ii^{n(2n+1)}\gamma^1\cdots\gamma^{2n} =
(-\ii)^n\gamma^1\cdots\gamma^{d=2n}\ ,
\eea
in the Euclidean signature, although this detail enters our discussion
explicitly only for the computation in the Appendix.

The regulated functional trace
\bea\label{AGW_true}
{\rm Tr}(\Gamma \hat{\boldsymbol\delta}_\xi) =\lim_{\beta\rightarrow 0}{\rm Tr}\left[\Gamma e^{-\beta((\ii\gamma^a\bD_a)^2- \beta^{-1}\hat{\boldsymbol\delta}_\xi)}\right]\bigg|_{\xi{\rm\textendash linear}}
\eea
can then be either computed by the Heat Kernel expansion or recast as
the Euclidean path integral of this supersymmetric quantum mechanics
with periodic boundary conditions on the fermions, as is well known.
We will take the former methodology for actual computation down the road.

A well-known fact is that the covariant anomaly in general is not  really the anomalous variation
of the effective action, also known as the consistent anomaly,
\bea
\hat{\boldsymbol\delta}_\xi W\; \neq \;{\rm Tr}(\Gamma \hat{\boldsymbol\delta}_\xi)\ .
\eea
$\hat{\boldsymbol\delta}_\xi W$ is  the
physical quantity that appears on the right hand side of the anomalous Ward identity.
However, also well known is that the anomaly polynomial we find from the right hand
side determines the left hand side  via a procedure called the anomaly descent,
\bea
\hat{\boldsymbol\delta}_\xi W\; \leftrightarrow \;{\rm Tr}(\Gamma \hat{\boldsymbol\delta}_\xi)\ .
\eea
 Only because
the cancelation of the covariant anomalies on the right implies the cancellation of the consistent ones
on the left, and vice versa, we usually do not strain to distinguish the two objects.

The effective action $W$ is entirely a functional of the metric (and the gauge
fields). Once we arrive at $W$,  no fermions reside there anymore, so there
is no need  for the Kosmann lift. The Wess-Zumino consistency
condition is far simpler in the coordinate basis, where the  effect of
$\hat \xi_K$ would be washed out entirely. This Wess-Zumino consistency conditions
for diffeomorphisms were addressed fully by Bardeen and Zumino \cite{Bardeen:1984pm},
who discovered that a naive anomaly descent
using only the rotational part of the diffeomorphism automatically solves the full consistency
condition that involves the translational parts as well. As noted already, however,
this standard process for the consistent anomaly does not address the overall normalization
and the anomaly polynomial, so the covariant anomaly enters here and fixes everything in
conjunction with the consistent side.

\subsection{The Legacy Computation}

As such, the place where the Kosmann lift should have entered was the computation
of the covariant diffeomorphism anomaly, or the functional traces such as (\ref{AGW_true}).
In Ref.~\cite{Alvarez-Gaume:1983ihn}, on
the other hand, the authors seemingly started with
\bea
\boldsymbol\delta_\xi'\Psi = \xi^\mu\partial_\mu\Psi
\eea
as a generator of the diffeomorphism, which we already argued against on the basis
of how it lacks a global definition. Have they computed the anomaly
\bea\label{AGW_false}
{\rm Tr}(\Gamma \boldsymbol\delta'_\xi) =\lim_{\beta\rightarrow 0}{\rm Tr}\left[\Gamma e^{-\beta((\ii\gamma^a\bD_a)^2- \beta^{-1}\boldsymbol\delta'_\xi)}\right]\bigg|_{\xi{\rm\textendash linear}}\ ,
\eea
under this illegal local coordinate transformation?

On a closer inspection, however, one can see that at some point of the computation
 $\boldsymbol\delta'_\xi$ is replaced by its covariantized version
\bea
\xi^\mu\partial_\mu \qquad\Rightarrow\qquad\xi^\mu {\bD}_\mu \ .
\eea
With the bad behavior of $  \boldsymbol\delta'_\xi$ beyond local patches,
there would have been no practical alternative, sans the Kosmann lift.
Furthermore, even though this is not stated explicitly in Ref.~\cite{Alvarez-Gaume:1983ihn},
the real computation proceeded with a further shifted operator
\bea
\boldsymbol\Delta_\xi \equiv \xi^\mu \bD_\mu+\frac12(\nabla_\mu\xi^\mu) \ ,
\eea
with the new additive piece $\nabla_\mu \xi^\mu/2$ if we rephrase it as a
functional trace.

The shift was implicit in the actual
computation by the time they recast the problem as a path
integral of certain supersymmetric quantum mechanics. When we connect this
path integral back to the canonical side, operators are naturally  normal-ordered in
the end, which translates to the replacement
\bea
\xi^\mu \bD_\mu\qquad\Rightarrow\qquad \frac12\left(\xi^\mu \bD_\mu +\bD_\mu\xi^\mu\right)= \xi^\mu \bD_\mu + \frac12(\nabla_\mu\xi^\mu)\ .
\eea
This essential shift was emphasized later by Fujikawa \cite{Fujikawa:1985}. On the
path integral side, this normal ordering can also
be understood from how the measure of the path integral must be built from the eigen-modes,
weighted by volume factor, $({\rm det}\,g)^{1/4}\,\psi_j$. The latter combination is
sometimes referred to as ``half-density" spinor \cite{Bastianelli}.

The same can be seen also easily from the operator realization
of the functional trace. It comes from how  operators sandwiched by eigenfunctions
would be expressed,
\bea
\langle\psi_j\vert \cO\vert \psi_k\rangle =\int d^d x\;\sqrt{g}\,\psi_j^\dagger \cO\psi_k
=\int d^d x\; ({g}^{1/4}\psi_j)^\dagger \cO'(g^{1/4}\psi_k)\ ,
\eea
where the precise realization of $\cO'$ as differential operator would differ
from how we usually view the abstract $\cO$ if the latter act on $g^{1/4}$ nontrivially.
In the above case, $\bD$ would annihilate $g^{1/4}$, yet the Lie derivative that
motivated $\xi^\mu \bD_\mu$ does affect the metric factor. One can thus see that
this line of thought results in the shift by $(\nabla_\mu\xi^\mu)/2$.

This series of substitutions resulted in
\bea\label{AGW_diffeo}
{\rm Tr}(\Gamma  \boldsymbol\Delta_\xi )&=&\lim_{\beta\rightarrow 0}
{\rm Tr}\left[\Gamma e^{-\beta((\ii\gamma^a\bD_a)^2- \beta^{-1} \boldsymbol\Delta_\xi)} \right]\bigg|_{\xi{\rm\textendash linear}}\cr\cr
&=& \int P_{d+2}(\IR_{\beta} ^{\;\;\;\alpha}  +2\pi\ii(- \nabla_\beta\xi^\alpha+\nabla^\alpha\xi_\beta);\cF ) \,\biggr\vert_{\hbox{\scriptsize $\xi$-linear}}\cr\cr
&=& \int P_{d+2}(\IR_{\beta} ^{\;\;\;\alpha}  +4\pi\ii(- \nabla_\beta\xi^\alpha);\cF ) \,\biggr\vert_{\hbox{\scriptsize $\xi$-linear}}\ ,
\eea
for some anomaly polynomial $P_{d+2}$ and the curvature 2-form,
\bea
\IR_{\beta}^{\;\;\;\alpha} = -\frac12 R^{\alpha}_{\;\;\beta\mu\nu}dx^\mu\wedge dx^\nu \ .
\eea
We use the antisymmetric nature of $\IR_{\beta\alpha}$ for the last step in (\ref{AGW_diffeo}).

Here we are using a slightly nonconventional form of the Riemann curvature that
has a close parallel to the Yang-Mill curvature, in that
\bea
\IR = d\mathbb{\Gamma} +{\mathbb\Gamma}\wedge {\mathbb\Gamma}\ ,
\eea
with the Christoffel symbol packaged into a connection 1-form \cite{Bardeen:1984pm}
\bea
\qquad
{\mathbb\Gamma}_{\beta}^{\;\;\alpha}\equiv -\Gamma^\alpha_{\;\;\mu\beta}\, dx^\mu\ .
\eea
Nominally, $\mathbb{\Gamma}$ is $GL(d)$-valued, even though, component-wise,
$\mathbb{R}$ is   the same old Riemann curvature, as noted above.

One might think that, for the pure gravitational cases, the difference between $\xi^\mu\partial_\mu$
and $\xi^\mu {\mathscr D}_\mu$ part of $\xi^\mu {\bD}_\mu$ is a local Lorentz transformation,
thus the substitution is harmless. However, since the very presence of the diffeomorphism
anomaly implies the anomalous local Lorentz transformation, the substitution
is hardly innocuous. Besides, the difference, $-\xi^\mu w_{\mu ab}\gamma^{ab}/4$,
is not even a valid gauge function.  It appears that
the classic result in Ref.~\cite{Alvarez-Gaume:1983ihn} is neither for the naive
${\boldsymbol\delta}'_\xi$ nor for the generalized Kosmann lift $\hat{\boldsymbol\delta}_\xi$.

\subsection{Generalized Kosmann Comes to the Rescue}

The difference between the generalized Kosmann lift ${\hat{\boldsymbol\delta}}_\xi$ on spinors
and the above $\xi^\mu {\mathscr D}_\mu$ looks more sensible, on the other hand, with
\bea
{\hat{\boldsymbol\delta}}_\xi \Psi - \xi^\mu {\bD}_\mu \Psi
%=\left({\hat{\boldsymbol\delta}}_\xi  - \xi^\mu {\mathscr D}_\mu \right) \Psi
= -\frac14 \hat\xi^{ab}_{V}\gamma_{ab}\Psi \ .
\eea
The difference is covariant and a special form of a local Lorentz transformation
with  $-\hat\xi^{ab}_{V}= D^{[a}\xi^{b]}$. Recall from the previous section how
this transformation rule ${\hat{\boldsymbol\delta}}_\xi$ with gauge fields included in $\mathscr D$
as well  preserves the action when it is invoked along with the
Kosmann lift $\mathscr L_\xi$ on the vielbein and on the spin connection. In retrospect, the covariant
diffeomorphism anomaly should have been computed with this operator ${\hat{\boldsymbol\delta}}_\xi$ as the generator.

As noted already, in performing the relevant functional trace, the same should be viewed as
\bea
{\hat{\boldsymbol\delta}}_\xi  (({\rm det}\,g)^{1/4}\Psi)-\boldsymbol\Delta_\xi( ({\rm det}\,g)^{1/4}\Psi)
%=\left({\hat{\boldsymbol\delta}}_\xi  - \xi^\mu {\mathscr D}_\mu \right) \Psi
= -\frac14 \hat\xi^{ab}_{V}\gamma_{ab}(({\rm det}\,g)^{1/4}\Psi)\ ,
\eea
where we kept the common notation $\hat{\boldsymbol\delta}_\xi$ for its natural generalization
to the ``half-density" spinors, with the relevant shift already evident via the same
shift in $\hat{\boldsymbol\delta}_\xi$.
The operator ${\boldsymbol\Delta}_\xi$ is therefore better understood as part of $\hat{\boldsymbol\delta}_\xi$,
or as a combination of the proper diffeomorphism that starts from $\mathscr L_\xi$ and acts on the
``half-density" spinor and an additional local Lorentz transformation of $\sim -\hat\xi_V$.

We must recompute the covariant diffeomorphism anomaly
\bea\label{with_Kosmann}
G^{\rm cov}_{\rm diff}(\xi)&=& \lim_{\beta\rightarrow 0}{\rm Tr}\left(\Gamma e^{-\beta \cQ}\,{\hat{\boldsymbol\delta}}_\xi\right)\cr\cr
& =& \lim_{\beta\rightarrow 0} {\rm Tr}\left(\Gamma e^{-\beta  \cQ}\left[ {\boldsymbol\Delta}_\xi  -\frac14 \hat\xi^{ab}_{V}\gamma_{ab}\right]\right)
\eea
from scratch. The first piece was already computed by Alvarez-Gaume and Witten, so it falls
upon us to compute the second, additional piece.
In view of how we advocated the generalized Kosmann
lift, the question comes down to how this additional piece
would have figured into the computation in Ref.~\cite{Alvarez-Gaume:1983ihn}.

For instance, the difference due to $-\hat\xi^{ab}_{V}\gamma_{ab}/4$ as in
\bea\label{additional}
{\rm Tr}\left[\Gamma e^{-\beta((\ii\gamma^a\bD_a)^2 + \beta^{-1} \hat\xi^{ab}_{V}\gamma_{ab}/4)} \right]\bigg|_{\xi{\rm\textendash linear}}
\eea
starts out at a negative power of $\beta$,
\bea
\sim \frac{1}{\beta}\int (d\xi)\wedge {\rm tr}\,(\,\cdots)\ ,
\eea
where $\xi$ is treated as 1-form and the ellipsis is a sum of wedge products of $(d/2-1)$-many
curvature 2-forms, $\cF$ and $\cR$. The Bianchi identities for the curvature 2-form
implies $d [{\rm tr}\,(\,\cdots)]=0$, so that this term vanishes upon integration by parts
on a compact spacetime or with vanishing asymptotic boundary condition on the curvatures.

The question is therefore what happens to $\beta^0$ terms that come about
due to $-\hat\xi^{ab}_{V}\gamma_{ab}/4$ part of $\hat{\boldsymbol\delta}_\xi$. This computation,
when attacked directly, is substantially more involved than the conventional computation
of the covariant  anomaly. In Appendix A, we will take a $d=4$ example and make such
 brute-force evaluations for an illustration of what kinds of computations are involved.

On the other hand, it turns out that there is a far simpler way to evaluate
(\ref{additional}), relying on a general property of the functional trace by noting the vanishing identity,
\bea\label{vanishing}
0={\rm Tr}\biggl(\Gamma e^{-\beta \cQ}\,\left[(\gamma_b\xi^b)(\gamma^a\bD_a)+(\gamma^a\bD_a)(\gamma_b\xi^b)\right]\biggr)\ ,
\eea
by virtue of $\Gamma\gamma_a+\gamma_a\Gamma=0$, $\cQ=-(\gamma^a\bD_a)^2$, and
the cyclic property of ${\rm Tr}$. As with any such formal argument, the last cyclic property
is something we need to be wary of in the functional setting; nevertheless, it will
hold, given sufficiently nice boundary conditions.
It follows immediately from $\{\gamma_b\xi^b, \gamma^a\bD_a\}= 2{\boldsymbol\Delta}_\xi  - \hat\xi^{ab}_{V}\gamma_{ab}$
that
\bea
\lim_{\beta\rightarrow 0}{\rm Tr}\left(\Gamma e^{-\beta \cQ}\,{\boldsymbol\Delta}_\xi\right) \;=\;
\frac12\times\lim_{\beta\rightarrow 0}{\rm Tr}\left(\Gamma e^{-\beta  \cQ}\, \hat\xi^{ab}_{V}\gamma_{ab}\right)\ ,
\eea
for the above functional traces, either on a compact manifold or with a physical boundary
condition that enforces a fast asymptotic vanishing of the field strengths.

The quantity of interest (\ref{with_Kosmann}) is a linear combination of the two sides
of this equality with the additional factors of 1 and $-1/2$, respectively,
\bea\label{key_diffeo}
\lim_{\beta\rightarrow 0}{\rm Tr}\left(\Gamma e^{-\beta \cQ}{\hat{\boldsymbol\delta}}_\xi \right)&=&
\lim_{\beta\rightarrow 0}{\rm Tr}\left(\Gamma e^{-\beta \cQ}\left[{\boldsymbol\Delta}_\xi  -\frac14 \hat\xi^{ab}_{V}\gamma_{ab}\right]\right) \ , %\cr\cr&=&\lim_{\beta\rightarrow 0}{\rm Tr}\left(\Gamma e^{-\beta \cQ}\hat{\boldsymbol\delta}_\xi \right)
\eea
which brings us immediately to
\bea\label{AGW+Kosmann}
G^{\rm cov}_{\rm diff}(\xi)&=& \lim_{\beta\rightarrow 0}{\rm Tr}\left(\Gamma e^{-\beta \cQ}\,{\hat{\boldsymbol\delta}}_\xi\right)
\;=\;\frac12\times
\lim_{\beta\rightarrow 0}{\rm Tr}\left(\Gamma e^{-\beta \cQ}\,{\boldsymbol\Delta}_\xi \right)\cr\cr
&=&
\int P_{d+2}(\IR_{\beta} ^{\;\;\alpha}  +2\pi\ii(- \nabla_\beta\xi^\alpha);\cF ) \,\biggr\vert_{\hbox{\scriptsize $\xi$-linear}}\ ,
\eea
in the end, with $2\pi\ii$ replacing $4\pi\ii$ of (\ref{AGW_diffeo}) but for precisely the same
old anomaly polynomial $P_{d+2}$.

It is important to note that none of these affect the anomaly polynomials, which
are the most important and most widely used results of Ref.~\cite{Alvarez-Gaume:1983ihn}.
These venerable anomaly polynomials stand uncorrected, despite the generalized Kosmann lift, or
in  retrospect,  more properly justified thanks to the generalized Kosmann lift.

\subsection{Kosmann for Other Chiral Fields}

The most general form of the anomaly polynomial is \cite{Alvarez-Gaume:1983ihn}
\bea
P_{d+2}(\mathbb{R},\cF)=\mathbb A(\cR)\boldsymbol\wedge_{{\mathbf r}} \ch_{\mathbf r}(\cF)\biggr\vert_{\hbox{\scriptsize $(d+2)$-form}}
\eea
with the A-roof genus $\mathbb{A}$ from the spinor bundle. On the right, we used
the curvature 2-form in its Cartan-Maurer form $\cR$, yet this can be easily
translated to $\mathbb{R}$ inside the traces.  The Chern
classes $\ch$'s arise from vector bundles in representations ${\mathbf r}$ of some gauge groups,
of which the chiral fields in question are also sections. The simplest context we
worked on so far assumed that these vector bundles are all associated with internal gauge
symmetries.

For more general chiral fields, we would find additional $\ch(\cR)$'s among the
latter factors.
Rarita-Schwinger field, $\Psi_a$, for example, carries an extra local Lorentz index
that is contracted with a Dirac matrix in the Lagrangian,
\bea
\sim\bar \Psi_a\gamma^{abc}\sD_b\Psi_c \ ,
\eea
so it is natural to take the local Lorentz index instead of the coordinate one. As such
after taking into account the usual ``traceless" condition $\gamma^a\Psi_a=0$, it
is natural to expect that the anomaly polynomial would arise from
\bea\label{RS_anomaly_P}
\mathbb A(\cR)\wedge \left(\ch_{\rm def}(\cR) -1\right)\boldsymbol\wedge_{{\mathbf r}} \ch_{\mathbf r}(\cF) \ .
\eea
Let us concentrate on this example and see how the Kosmann lift again enters the
covariant anomaly computation and produces an answer on par with the spinor case of the
previous subsection.

An immediate question is how the Kosmann-lifted Lie derivative acts on such a higher-spin
chiral field. Starting from the earlier observation in (\ref{Kosmann_v}),
\bea
\mathscr{L}_\xi v^a\;=\; \xi^c D_c v^a - \hat\xi^{ac}_{V} v_c
\eea
and using the Leibniz rule, we arrive at
\bea
\mathscr{L}_\xi u_b\;=\; \xi^c D_c u_b + u^c \hat\xi_{Vcb}  \;=\; \xi^c D_c u_b - \hat\xi_{Vbc} u^c \ .
\eea
This means that on the Rarita-Schwinger field, the (generalized) Kosmann acts as
\bea
\hat{\boldsymbol\delta}_\xi \Psi_a =\xi^\mu\bD_\mu  \Psi_a
+ {\boldsymbol\delta}^{\rm local\; Lorentz}_{\hat\xi_V} \Psi_a =\left(\xi^\mu\bD_\mu -\frac14 \hat\xi^{bc}_{V}\gamma_{bc}\right)  \Psi_a
- \hat\xi_{Va}^{\;\;\;\ b} \Psi_b
\eea
where $\bD$ now includes the spin connection acting on the 1-form orthonormal
index of $\Psi_a$ as well.

Once we write out the translation operator this way, it is clear that
$\hat{\boldsymbol\delta}_\xi$ acting on the Rarita-Schwinger field
is equipped with the additional local Lorentz rotation on the
lower orthonormal index of $\Psi_a$ by the amount of $\Theta=\hat\xi_V$.
We remind the readers of the convention (\ref{gauge_convention}) in use
for gauge transformations. The covariant anomaly computation due
to this additional part of the translation operator would then proceed
exactly the same way as with other internal gauge sectors,\footnote{We skip
this more straightforward part of the computation, which originates from Ref.~\cite{Alvarez-Gaume:1983ihn},
closely modeled after the Fujikawa computation of the axial anomaly.
We quote the same for the discussion in the next subsection, as well.} so that the shift of the curvature
\bea
\cR_{ab}\quad\rightarrow\quad \cR_{ab}+ 2\pi\ii \, \hat\xi_{Vab}
\eea
occurs inside $\ch_{\rm def}(\cR)$ when the latter contributes to the covariant anomaly.

Using $\hat\xi_{Vab}= - D_{[a}\xi_{b]}= (- D_{a}\xi_{b} + D_{b}\xi_{a})/2  $,
and translating back to the coordinate basis, this effectively gives
\bea
\mathbb{R}_\alpha^{\;\;\beta} \quad\rightarrow\quad \mathbb{R}_\alpha^{\;\;\beta} + 2\pi\ii \, (-\nabla_\alpha \xi^\beta)
\eea
for all instances of the Riemann curvature 2-forms in the anomaly polynomial.
This allows
\bea\label{cov_general}
G^{\rm cov}_{\rm diff}(\xi)&=&
\int P_{d+2}(\IR_{\beta} ^{\;\;\alpha}  +2\pi\ii(- \nabla_\beta\xi^\alpha);\cF ) \,\biggr\vert_{\hbox{\scriptsize $\xi$-linear}}\ ,
\eea
for Rarita-Schwinger field as well, with $P_{d+2}$ computed from (\ref{RS_anomaly_P}).

For (anti-)chiral tensors in $d=4k+2$, say, $B$ and $H=dB+\cdots$, the computation of the covariant anomaly should
proceed similarly. In Ref.~\cite{Alvarez-Gaume:1983ihn}, the problem was recast by
starting from a field constructed from a tensor product of a spinor and an additional
Weyl spinor. Although this tensor product is equivalent to not only $B$ and $H$ but
includes other differential forms, the latter are not chiral and thus deemed harmless
for the anomaly computation. This means that all that changes is  again  a matter of
an extra factor in the relevant characteristic class,
\bea
-\,\mathbb A(\cR)\wedge \frac12\,\ch_{\rm Weyl\;spinor}(\cR)\boldsymbol\wedge_{{\mathbf r}} \ch_{\mathbf r}(\cF)
\eea
where the half originates from the reality of $B$ and $H$ and the overall
sign from the bosonic nature of these chiral fields. The additional Weyl spinor index acts
much like a gauge index, and is rotated, by virtue of the Kosmann lift, under a local
Lorentz gauge transformation of $\Theta=\hat\xi_V$. This, combined with the spinor
computation of the previous subsection, again leads us to the universal
shift of the curvature as in (\ref{cov_general}).

One should note  how the necessary shift of the Riemann curvature in these additional
$\ch(\cR)$ contributions emerges  entirely from the
Kosmann lift. In other words,  when it comes to the anomaly contribution from these $\ch(\cR)$ due to
higher spin content, it is not that the Kosmann lift halves the naive expectations
from $\boldsymbol\Delta_\xi$, but oppositely, the necessary shift that computes the covariant anomaly
would occur only if one chooses to invoke the Kosmann lift.

Conventionally, on the other hand, such a shift of the curvature 2-form
was engineered by instituting a rotation of the additional index, e.g., 1-form index
of $\Psi_\mu$, by $\nabla_\mu\xi_\nu-\nabla_\nu\xi_\mu$. Note how the latter
differs, again, from the effect encoded in the vanilla Lie derivative $\mathfrak L_\xi$ on $\Psi_\mu$.
On par with the case of spinors, this choice again incurs an additional factor 2, relative to the
Kosmann, so in this sense, we again find that the same factor 1/2 reduction relative to the
result of Ref.~\cite{Alvarez-Gaume:1983ihn} is necessary for higher spin chiral fields as well.

\subsection{Consistent from Covariant}

In the past,  such a factor 1/2 has been noticed: Ref.~\cite{Bastianelli}
discusses various alternative translational operators, among which are $\delta_{cov}$ and $\delta_{sym}$.
In our notations, the former equals $\xi^\mu\bD_\mu$, while the latter was motivated as a combination
of $\delta_{cov}$ and a local Lorentz rotation so that its action on the vielbein produces a naturally
symmetric energy-momentum tensor $T$. Recalling the general discussion in the header of Section 5, it
should be obvious this latter requirement would enforce $\delta_{sym}=\mathscr L_\xi$ on the vielbein, so
the computation that starts from  $\delta_{sym}$ would eventually lead to a covariant anomaly with the same
factor 1/2 relative to that of $\delta_{AGW}$, again in their notation.

What remained unclear was whether these various choices, regarded as different
combinations of diffeomorphisms and local Lorentz rotations, are merely equivalent
representations of the same physical phenomenon;
as long as one is interested in cancelation of the gravitational anomalies,
such a universal factor 1/2 depending on the precise operator chosen, may seem
innocuous. However, what really appears in physical Ward identity and in the anomalous effective
action is the consistent anomaly, the normalization of which is a serious matter since
in string theory, for example, such anomalies are often canceled by inflows from some definite
topological couplings.

In particular, the consistent anomaly \cite{Wezz:1971,Zumino:1983,Stora:1984}
for diffeomorphism is compactly written in the
coordinate basis with the curvature 2-form $\IR$ \cite{Bardeen:1984pm},  with the local Lorentz index
absent. This means that for this side, there is no ambiguity from the local Lorentz
symmetry mixing in, and there has to be a unique answer that sits on the right side
of the diffeomorphism Ward identity. The covariant anomaly should be considered
only as an intermediate step toward the consistent anomaly, on the other hand, which
supplies the anomaly polynomial and the normalization,  so an unambiguous question
emerges to ask which version of the various purported covariant anomalies makes
sense by the time we reached the consistent side.

The general connection between the two sides is originally from Ref.~\cite{Bardeen:1984pm},
both for gauge anomalies and their diffeomorphism cousin, although it did not fix
this crucial normalization for the latter we seek here.
With our unambiguous motivation in favor of the Kosmann lift, we expect that the
covariant anomaly (\ref{cov_general}) we computed should lead to the
generally anticipated consistent anomaly with the right coefficient.
In this last part, we will revisit consistent diffeomorphism anomaly and show that the numerical
factor we found for the covariant side under the Kosmann transfers to the consistent side and how the resulting
normalization fits precisely the general anomaly descent and thus the anomaly inflow mechanism.

Let us start by recalling how this went for gauge anomalies which are generally less cumbersome.
The anomaly polynomial $P_{d+2}(\cF)$ with gauge field strength 2-forms $\cF$
is determined in the process of computing the covariant anomaly
\bea
G^{\rm cov}= \int P_{d+2}(\cF+2\pi\ii\Theta)\biggr\vert_{\Theta\text{-linear}}
\eea
with the gauge function $\Theta$ \cite{Alvarez-Gaume:1983ihn}. The normalization of $P_{d+2}$ itself
is tied to $(d+2)$-dimensional Atiyah-Singer index formulae. The anomaly descent, with the same $P_{d+2}(\cF)$,
\bea
 P_{d+2}(\cF)=d\bw_{d+1}^{(0)}(\cA,\cF)\ , \qquad   {\boldsymbol\delta}_{\Theta}\bw_{d+1}^{(0)}(\cA,\cF)
 =d\bw_d^{(1)}(\Theta; \cA,\cF)\ ,
\eea
isolates the consistent anomaly,
\bea
G\equiv {\boldsymbol\delta}_{\Theta}W(\cA)= 2\pi\ii\, \int \bw_d^{(1)}(\Theta;\cA,\cF)\ ,
\eea
with $2\pi \ii$  inherited from that of the covariant anomaly, on the other hand.
The two sides famously reconcile as \cite{Bardeen:1984pm}
\bea\label{G2G_BZ}
G^{\rm cov}(\Theta;\cF) =G(\Theta;\cA,\cF) + \,d_\cA{\Theta} \circ (- 2\pi \ii \cK)
\eea
via the introduction of the Bardeen-Zumino current $\cK$, also determined by the same $P_{d+2}(\cF)$.

Here we shall accept these well-known facts and adapt the spirit to the case of diffeomorphism.
When we turn to the diffeomorphism anomaly, many of these conclusions
survive in the end, despite how various middle steps cannot be the same given that
the transformation involved is primarily translational,
\bea\label{diffeo_Gamma}
{\boldsymbol\delta}_\xi {\mathbb\Gamma} &=&\mathfrak{L}'_\xi{\mathbb\Gamma}+ {{\boldsymbol\delta}}_{-\partial\xi}^{GL(d)}\mathbb{\Gamma}\ , \qquad{{\boldsymbol\delta}}_{-\partial\xi}^{GL(d)}\mathbb{\Gamma} \equiv { d}_{\mathbb\Gamma} (-\partial \xi)\ ,
\eea
where $-\partial \xi$ represents a $GL(d)$ matrix, $-\partial_\beta\xi^\alpha$, playing the
role of the gauge function $\Theta$.
Here, ${\mathfrak L}'_\xi$ treats ${\mathbb\Gamma}_{\beta}^{\;\;\alpha}$ as if
it is a collection of 1-forms and ignores the other $GL(d)$ indices, $\alpha$ and $\beta$.
The ``gauge"
transformation of the $GL(d)$ connection ${\mathbb\Gamma}_{\beta}^{\;\;\alpha}$ occurs almost as an
afterthought.

A nontrivial fact is that despite this very distinct action of the diffeomorphism,
the consistent anomaly is given by the naive $GL(d)$ anomaly descent \cite{Bardeen:1984pm},
\bea
G_{\rm diff}(\xi;{\mathbb\Gamma},\IR;\cF)\equiv {\boldsymbol\delta}_{\xi}W(\mathbb{\Gamma};\cA)= 2\pi\ii\, \int \bw_d^{(1)}(-\partial\xi;\mathbb{\Gamma},\mathbb{R};\cF)
\eea
as if the  translational action $\mathfrak{L}'_\xi$ was absent.
Although we wrote $2\pi\ii$ as the multiplicative coefficient in line with
general anomaly descent, the consistent side is incapable of determining
either this overall coefficient or even the anomaly polynomial. The aim
is to see that this $2\pi\ii$ is precisely inherited from the same in the
covariant side (\ref{cov_general}).

In other words, we wish to show that, with a Bardeen-Zumino current  $\mathbb{K}$
to be shown explicitly below,
\bea\label{diffeo_cov2consistent}
G^{\rm cov}_{\rm diff}(\xi)\;=\; G_{\rm diff}(\xi;{\mathbb\Gamma},\IR;\cF) +{\boldsymbol\delta}_\xi\mathbb{\Gamma}\circ (-2\pi\ii \mathbb{K})\ ,
\eea
in the general manner as in (\ref{G2G_BZ}), with aforementioned  $G_{\rm diff}$ and
\bea
G^{\rm cov}_{\rm diff}(\xi)%&=& \lim_{\beta\rightarrow 0}{\rm Tr}\left(\Gamma e^{-\beta \cQ}\,{{\boldsymbol\delta}}_\xi\right)\cr\cr
&=&
\int P_{d+2}(\IR_{\beta} ^{\;\;\alpha}  +2\pi\ii(- \nabla_\beta\xi^\alpha);\cF ) \,\biggr\vert_{\hbox{\scriptsize $\xi$-linear}}\ .
\eea
All quantities, including  $\mathbb{K}$,  are
derived from a common anomaly polynomial $P_{d+2}(\mathbb{R};\cF)$.

The additional factor $1/2$, thanks to the generalized Kosmann lift, is essential for the matching
down to numbers; we will take time to trace through the relation between the two versions of the
diffeomorphism anomalies. In the end we will find, the Bardeen-Zumino current here is
again a pure gauge-type, now with respect to the naive $GL(d)$, and can be expressed as
\bea
\mathbb{K}=\frac{\boldsymbol\partial}{\boldsymbol\partial \mathbb{R}}\bw_{d+1}^{(0)}(\mathbb{\Gamma}, \mathbb{R};\cF)\ ,
\eea
where $d\bw_{d+1}^{(0)}(\mathbb{\Gamma}, \mathbb{R};\cF)=P_{d+2}(\mathbb{R};\cF)-P_{d+2}(0;\cF) $
as is familiar from the anomaly descent. In turn, the same $\bw_{d+1}^{(0)}$ produces the consistent anomaly above
via
\bea
\bw_{d}^{(1)}(\Omega;{\mathbb\Gamma},\IR;\cF) \equiv {\rm tr}\left(\Omega\,\frac{\boldsymbol\partial}{\boldsymbol\partial \mathbb{\Gamma}}\bw_{d+1}^{(0)}(\mathbb{\Gamma}, \mathbb{R};\cF )\right)\ ,
\eea
for an arbitrary $GL(d)$-valued $\Omega$, equivalent to the usual
descent mechanism along the diffeomorphism side.

A key that allows us to do this is how $-\nabla_\alpha\xi^\beta$ that enters
$G^{\rm cov}_{\rm diff}(\xi)$ but differs from its counterpart $-\partial_\alpha\xi^\beta$ for
the anomaly descent, may be recast as
\bea
-\nabla_\alpha\xi^\beta =-\partial_\alpha\xi^\beta +\xi\lrcorner\,\mathbb{\Gamma}_\alpha^{\;\;\beta} \ ,
\eea
with $\mathbb{\Gamma}$ considered as a matrix-valued 1-form. Starting with this and after some
manipulations using the anomaly descent algebra, we may rewrite $G^{\rm cov}_{\rm diff}(\xi)$
as
\bea\label{diffeo_unwanted}
G^{\rm cov}_{\rm diff}(\xi) &=&G_{\rm diff}(\xi;{\mathbb\Gamma},\IR;\cF )  + \left(\mathfrak L_\xi'\mathbb{\Gamma}+ d_\mathbb{\Gamma} (-\partial\xi)\right)\circ (-2\pi\ii \mathbb{K})
\\\cr
&&+ \,2\pi\ii\int \bw_{d}^{(1)}(\xi\lrcorner\,\mathbb{\Gamma}; \mathbb{\Gamma},\mathbb{R};\cF ) +\,2\pi\ii \left(\mathfrak L_\xi'\mathbb{\Gamma}-d_\mathbb{\Gamma} \left( \xi\lrcorner\,\mathbb{\Gamma}\right)\right)\circ \mathbb{K}\ , \nonumber
\eea
where we used (\ref{G2G_BZ}) strictly for $GL(d)$ and the claimed form of $G_{\rm diff}$ twice, once
with $\Theta\rightarrow -\partial\xi$ and one more time with  $\Theta\rightarrow \xi\lrcorner\,\mathbb{\Gamma}$.

The first line on the right of (\ref{diffeo_unwanted}) is precisely the desired right hand side of (\ref{diffeo_cov2consistent}),
so the task boils down to how the remainders in the second line cancel out among themselves. With the identity
\bea
\left(\mathfrak L_\xi'\mathbb{\Gamma}-d_\mathbb{\Gamma} \left( \xi\lrcorner\,\mathbb{\Gamma}\right)\right)\!{}_{\alpha\;\;\;\mu}^{\;\;\,\beta}
%&=&\xi^\lambda\partial_\lambda\mathbb{\Gamma}_{\mu\,\alpha}^{\;\;\;\;\;\beta} +(\partial_\mu\xi^\lambda)\mathbb{\Gamma}_{\lambda\,\alpha}^{\;\;\;\;\;\beta}\cr\cr
%&&-\left(\partial_\mu (\xi^\lambda\mathbb{\Gamma}_{\lambda\,\alpha}^{\;\;\;\;\;\beta})+\mathbb{\Gamma}_{\mu\,\alpha}^{\;\;\;\;\;\gamma} (\xi^\lambda\mathbb{\Gamma}_{\lambda\,\gamma}^{\;\;\;\;\;\beta}) -(\xi^\lambda\mathbb{\Gamma}_{\lambda\,\alpha}^{\;\;\;\;\;\gamma}) \mathbb{\Gamma}_{\mu\,\gamma}^{\;\;\;\;\;\beta} \right)\cr\cr
&=&\xi^\lambda\, \mathbb{R}_{\alpha\;\;\;\lambda\mu}^{\;\;\,\beta}\ ,
\eea
the unwanted second line of (\ref{diffeo_unwanted}) organizes into
\bea\label{diffeo_leftover}
2\pi\ii\int \bw_{d}^{(1)}(\xi\lrcorner\,\mathbb{\Gamma}; \mathbb{\Gamma},\mathbb{R};\cF )
+ 2\pi\ii\,(\xi\lrcorner\, \mathbb{R})\circ \mathbb{K} =2\pi\ii \int \xi\lrcorner' \,\mathbf{w}_{d+1}^{(0)}(\mathbb{\Gamma}, \mathbb{R};\cF)\ ,
\eea
where the contraction  $\lrcorner'$ is understood to be limited to $\mathbb{\Gamma}$ and $\mathbb{R}$.

Although $\mathbf{w}_{d+1}^{(0)}$ makes frequent appearances in the
descent mechanism, it does so only as mathematical middle steps. If one tries to
evaluate it with $d$-dimensional connections and curvatures inserted,
$\mathbf{w}_{d+1}^{(0)}=0$ identically since it is a $(d+1)$-form in the $d$-dimensional spacetime. The
vector $\xi$ is also  $d$-dimensional, so the contraction against it does
not affect the fact that (\ref{diffeo_leftover})
vanishes identically.
This concludes the demonstration of how the consistent anomaly is connected to the covariant one
in the usual manner, (\ref{diffeo_cov2consistent}).

In particular, the numerical factor $2\pi\ii$ of the covariant diffeomorphism
anomaly is inherited  by the consistent one. With $4\pi\ii$ in place of $2\pi\ii$, say,
for the chiral spinor contribution as in (\ref{AGW_diffeo}), one would have found an odd situation where the descent
of diffeomorphism comes with a factor 2 larger coefficient than other gauge
symmetries. In particular, the same factor 2 would have entered between the local
Lorentz anomaly and the diffeomorphism anomaly. This  odd situation is happily
avoided, again thanks to the generalized Kosmann lift.

\section{Summary}

We have explored various corners of the energy-momentum tensors and the Ward identities
thereof, with the Lie derivative proving to be the centerpiece of all these investigations.

The first half of the note is devoted to the question of the Noether procedure itself,
starting from an age-old statement that the naive procedure must be augmented by the
improvement term to agree with the symmetric energy-momentum tensor. There are
also literatures that  ``prove" how the two agree with each other, in contrast,
sometimes leaving us bewildered. We offer a simple and universal view on the
matter which shows how the Noether energy-momentum and the symmetric
energy-momentum are two sides of a single coin, so to speak.

The guiding principle is how, once coupled to the metric, the Lagrangian is generally covariant, meaning
that it is preserved modulo a total derivative under general coordinate
transformation. This in turn dictates that the position-dependent variation
of the matter field must follow the Lie derivative.
With this, the Lagrangian $d$-form density transforms universally by an exact $d$-form.
The vanishing rest is split into two mutually canceling parts, on the other hand,
one from the matter variation and the other from the spacetime variation. Each of these
produces the above two types of the energy-momentum tensor, respectively, and the mutual cancelation
demands the verbatim equality of the two types of energy-momentum tensors $T=\hat\iT$
in the end.

A key fact of life to note here is that the Lie derivative is relevant in flat
geometry as well, since even for simple translations the Lie derivative is unavoidable
when these are expressed in a curvilinear coordinate.  The Lie derivative per se has little to do with the
spacetime curvature as it is a fundamental structure of all differentiable manifolds
and exists prior to the introduction of the metric and the Levi-Civita connection.

This naturally brings us to the question of what should be the action of the diffeomorphism
on spinors. The prevalent answer to this in the physics community, which treats
spinors as if they consist of multiple scalars, is deficient if we recall how one
way to motivate the Lie derivative is as a directional derivative that maps a tensor to
a tensor. Unlike with tensors which can be defined once the manifold is equipped with
a differential structure, spinors are inevitably tied to the frame bundle, itself built
upon the vielbein and thus upon the metric, and equipped with the spin connection naturally.
Even though spinors are not  tensors in the latter's most restricted sense, a sensible
definition of a Lie derivative on the spinor must map it to another spinor, or a well-defined
section of the spin bundle. This eventually leads us to the Kosmann lift of the diffeomorphism,
which involves the covariant derivatives not just partial derivatives.

Equipped with this Kosmann lift of the diffeomorphism,
we went back to the matter of the energy-momentum tensor of spinors, where we show
that a mutually agreeing pair $T=\hat \iT$ again emerges in a naturally symmetric form, without resorting to the
equation of motion. Interestingly, we find that the  form of $T=\hat \iT$ is
more robust than the Lagrangian; without further tweaking, the same energy-momentum tensor
results from two different versions of the Dirac fermion, the canonical version
$\cL_{\rm Dirac}$ and the democratic version $\cL'_{\rm Dirac}$.

We must comment here that observations related to some of the above have appeared
in past literatures, a few of them quite recent. For instance, there is known ``proof" of how the Noether
energy-momentum tensor equals the symmetric one, which comes about from fixing
the ambiguous Noether procedure by demanding the internal gauge invariance \cite{Haberzettl:2024rai}.
This differs from ours in that we demand the general covariance which is far more
universal. Closer in spirit to ours is Ref.~\cite{Saravi:2002}, as noted already,
where the general covariance and the Lie derivative were used to demonstrate the
equality of the two energy-momentum for the Maxwell theory.

Another related work can be found in Ref.~\cite{Freese:2022}, which
recovered the correct energy-momentum tensor $\iT$ by taking into account the rotating
part of the Lie derivative, just as we did, although in detail the procedure is different.
In particular, the latter employs the variation of the spacetime integration measure
for the Noether procedure, often found in old physics literature on the subject.
This last goes against our general spirit that $\hat\iT$ should arise entirely
from the variation of the matter fields.

One can also find discussions of the
Kosmann lift for the computation of  $\iT$ \cite{Freese:2022,Helfer:2016,Bilyalov:1996} for Dirac fermions,
if not of why the end result must always equal to the symmetric energy-momentum tensor $T$.
We doubt that these few exhaust the relevant literature but
at least they show how unsettled the subject matter has been for many long years.
Finally, we introduced the notion of the generalized Kosmann lift, relevant when the
spinor in question is also coupled to gauge fields as well as to the spin connection.

All of these brought us to the matter of the diffeomorphism Ward identity.
Because this (generalized) Kosmann lift differs from the conventional choice, the existing
computation of the diffeomorphism anomaly must be rethought from scratch, which
we take up in Section 6. After recalling and clarifying a few subtleties
with the existing computations, such as how the translational operator used for
the computation was neither the Kosmann nor the naive version of the diffeomorphism
generator, we showed how the gravitation anomaly polynomials are unaffected
while the dictionary that extracts the anomaly from the latter must be modified
rather simply by an additional factor of $1/2$.

The legacy computation \cite{Alvarez-Gaume:1983ihn} should be regarded as a sum of two anomalies, in retrospect,
a violation of Kosmann-lifted diffeomorphism generated by $\xi$
and a violation of a local Lorentz transformation by the amount $-\hat\xi_V$. The two
happened to contribute equally to the covariant anomaly, coincidentally, and
in effect doubled the answer for the former. A factor 1/2
reduction is needed, therefore, for the pure diffeomorphism anomaly.
We also delineated how this factor $1/2$ propagates to the consistent
side and allows us to put the routine for the consistent diffeomorphism anomaly
on equal footing with gauge anomalies.\footnote{Although
we could not scan the extensive literature on the consistent side which
is by itself incapable of fixing the normalization,
it is very likely that the proper normalization has been in use for decades,
unknowingly, motivated by the simpler gauge anomaly side.}

In a sense, one of the more important ramifications of the later part of this note is
how the existing anomaly polynomials are now on completely solid ground, beyond
the precarious middle steps in the past, thanks to the Kosmann lift.
In turn, all these nooks and crannies emphasize strongly the importance of
the Lie derivative for quantum fields in general,  and, in particular, how the Kosmann-lifted Lie derivative
is not as an optional choice but rather a necessary part of the physics dictionary
for fermions.

\vskip 1cm

\section*{Acknowledgments}
We thank Sangmin Lee, Kentaro Hori, Yu Nakayama, and Alessandro Tomasiello for the discussion on various
aspects of this manuscript. This work  is supported by a KIAS individual grant, PG005705.

\appendix
\section{An Explicit Computation of $d=4$ Anomaly}

In this appendix, we will perform an explicit demonstration of the key identity
(\ref{key_diffeo}) for the case of $d=4$.
We will borrow heavily from Ref.~\cite{Barvinsky:1985} for the necessary
heat kernel expansion.
The simple derivation thanks to (\ref{vanishing}) is nice and powerful,
yet it fails to convey the nontrivial gymnastics underlying the equality.
We offer the appendix for a more explicit demonstration of the identity
(\ref{vanishing}) in favor of a better feeling of how things  work out in detail.

The Kosmann-lift contribution to the covariant diffeomorphism anomaly is
\bea\label{Kosmann}
\text{Tr}\left[\Gamma\left(-\frac14\hat\xi_V^{ab}\gamma_{ab}\right)\right]\,=\,\lim_{\beta\rightarrow 0}\text{Tr}\left[\Gamma\left(-\frac14\hat\xi_V^{ab}\gamma_{ab}\right)e^{-\beta(\ii\gamma^a\bD_a)^2}\right] \ .
\eea
As usual, we use the squared Dirac operator as a regulator.
\bea
(\ii\gamma^a\bD_a)^2=-\bD^2+\frac14R-\frac12\cF_{ab}\gamma^{ab}  \ .
\eea
Here, we will need the explicit form of the heat kernel in the coincidence limit,
\bea
\langle x|e^{-\beta(\ii\gamma^a\bD_a)^2} |x\rangle = \frac{\sqrt{g(x)}}{(4\pi\beta)^{d/2}}
\sum_{n=0}^\infty a_n(x)\beta^n\ ,
\eea
for which we read off some relevant coefficients from Ref.~\cite{Barvinsky:1985},
\bea
a_0\!&=&\!1 \ ,
\cr\cr
a_1\!&=&\!-\frac{1}{12}R+\frac12 \cF_{\mu\nu}\gamma^{\mu}\gamma^{\nu} \ ,
\cr\cr
a_2\!&=&\!
\frac{1}{180}R^{\mu\nu\rho\sigma}R_{\mu\nu\rho\sigma}-\frac{1}{180}R^{\mu\nu}R_{\mu\nu}+\frac{1}{288}R^2-\frac{1}{120}D^2 R+\frac{1}{12}\cF^{\mu\nu}\cF_{\mu\nu}
\cr\cr
&&+\left(\frac{1}{24}R\indices{^\kappa^\lambda_\mu_\nu}\cF_{\kappa\lambda}-\frac{1}{24}R\,\cF_{\mu\nu}+\frac{1}{12}D^2\cF_{\mu\nu}\right)\gamma^\mu\gamma^\nu
\cr\cr
&&+\left(\frac{1}{8}\cF_{\mu\nu}\cF_{\rho\sigma}+\frac{1}{192}R\indices{^\kappa^\lambda_\mu_\nu}R_{\kappa\lambda\rho\sigma}
\right)\gamma^\mu\gamma^\nu\gamma^\rho\gamma^\sigma \ .
\eea
These suffice for $d=4$.

Let us compute (\ref{Kosmann}) for $d=4$ using the above formulae, step by step.
Since we need the $\beta\rightarrow 0$ limit, it suffices to examine only $\beta^{-2}, \beta^{-1}, \beta^0$-terms.
First, it is evident that $\beta^{-2}$-term vanishes,
\bea
(\beta^{-2}\text{-term})\ =\ \int d^4 x\ \frac{\sqrt g}{(4\pi)^2\beta^2}\ \text{tr}\left[\Gamma\left(-\frac14\hat\xi_V^{ab}\gamma_{ab}\right)a_0\right]\, =\ 0\ ,
\eea
from the trace over spin indices, as usual.

Next up are $\beta^{-1}$-terms, for which we use  the usual relation
\bea
\text{tr}(\Gamma\gamma^\alpha\gamma^\beta\gamma^\mu\gamma^\nu)=\frac{-4}{\sqrt g}\epsilon^{\alpha\beta\mu\nu}
\eea
that we often invoke for computing Atiyah-Singer index densities, and find
\bea
(\beta^{-1}\text{-term})
\!&=&\!\int d^4 x\ \frac{\sqrt g}{(4\pi)^2\beta}\ \text{tr}\left[\Gamma\left(-\frac14\hat\xi_V^{ab}\gamma_{ab}\right)a_1\right]
\cr\cr
%&=&\frac{1}{128\pi^2}\int d^4 x\,\sqrt g\ D_{[\mu}\xi_{\nu]}\,\text{tr}(\cF_{\rho\sigma})\,
%\text{tr}(\Gamma\gamma^\mu\gamma^\nu\gamma^\rho\gamma^\sigma)
%\cr\cr
\!&=&\!\frac{-1}{32\pi^2\beta}\int d^4x \ D_{\alpha}\xi_{\beta}\,\text{tr}(\cF_{\mu\nu})\,\epsilon^{\alpha\beta\mu\nu} \ .
\eea
Integrating by parts, we find
\bea
(\beta^{-1}\text{-term})
\!&=&\!\frac{1}{32\pi^2\beta}\int d^4x \ \xi_{\beta}\,\text{tr}(D_{\alpha}\cF_{\mu\nu})\,\epsilon^{\alpha\beta\mu\nu}\ =\ 0 \ ,
\eea
by virtue of the Bianchi identity for $\cF$, as was claimed in the main text.

The last and the most involved $\beta^0$-terms are
\bea
(\beta^{0}\text{-term})
\!&=&\!\int d^4 x\ \frac{ \sqrt g}{(4\pi)^2}\ \text{tr}\left[\Gamma\left(-\frac14\hat\xi_V^{ab}\gamma_{ab}\right)a_2\right]
\cr\cr
\!&=&\!\frac{1}{16\pi^2}\int d^4 x\ D_{[\alpha}\xi_{\beta]}
\left[
\left(\frac{1}{4}\,\text{tr}(\cF_{\mu\nu}\cF_{\rho\sigma})
+\frac{1}{96}R\indices{^\kappa^\lambda_\mu_\nu}R_{\kappa\lambda\rho\sigma}\right)
g^{\alpha\mu}\epsilon^{\beta\nu\rho\sigma}\right.
\cr\cr
&&\left.+\frac{1}{24}
\bigg(\!
-R\indices{^\kappa^\lambda_\mu_\nu}\,\text{tr}(\cF_{\kappa\lambda})
+R\,\text{tr}(\cF_{\mu\nu})-2D^2\,\text{tr}(\cF_{\mu\nu})
\bigg)
\epsilon^{\alpha\beta\mu\nu}
\right]\ ,
\eea
from the following spinor trace formula
\bea
\text{tr}(\Gamma\gamma^\alpha\gamma^\beta\gamma^\mu\gamma^\nu\gamma^\rho\gamma^\sigma)
\,=\,\frac{-4}{\sqrt g}\left(g^{\alpha\beta}\epsilon^{\mu\nu\rho\sigma}
-g^{\alpha\mu}\epsilon^{\beta\nu\rho\sigma}
+g^{\beta\mu}\epsilon^{\alpha\nu\rho\sigma}\right.
\nonumber\\[1ex]
\left.+g^{\nu\rho}\epsilon^{\alpha\beta\mu\sigma}
-g^{\nu\sigma}\epsilon^{\alpha\beta\mu\rho}
+g^{\rho\sigma}\epsilon^{\alpha\beta\mu\nu}
\right)\ .
\eea
The first half  vanishes nontrivially with the help of another identity,
\bea
&&\left(\frac{1}{4}\,\text{tr}(\cF_{\mu\nu}\cF_{\rho\sigma})
+\frac{1}{96}R\indices{^\kappa^\lambda_\mu_\nu}R_{\kappa\lambda\rho\sigma}\right)
g^{\alpha\mu}\epsilon^{\beta\nu\rho\sigma}
\cr\cr
&&=\,\left(\frac{1}{4}\,\text{tr}(\cF_{\mu\nu}\cF_{\rho\sigma})
+\frac{1}{96}R\indices{^\kappa^\lambda_\mu_\nu}R_{\kappa\lambda\rho\sigma}\right)
\frac{1}{4}\,g^{\alpha\beta}\epsilon^{\mu\nu\rho\sigma}\ ,
\eea
checked  by brute-force.

Of the surviving $\beta^0$-terms
\bea
\frac{1}{384\pi^2}\int d^4 x\ D_{\alpha}\xi_{\beta}
\bigg(\!
-R\indices{^\kappa^\lambda_\mu_\nu}\,\text{tr}(\cF_{\kappa\lambda})
+R\,\text{tr}(\cF_{\mu\nu})-2D^2\,\text{tr}(\cF_{\mu\nu})
\bigg)
\epsilon^{\alpha\beta\mu\nu} \ ,
\eea
the third  can be manipulated via the Bianchi identity to
\bea\label{third piece}
D^2\,\text{tr}(\cF_{\mu\nu})\,\epsilon^{\alpha\beta\mu\nu}
\!&=&\!2D^\kappa D_\mu \,\text{tr}(\cF_{\kappa\nu})\,\epsilon^{\alpha\beta\mu\nu}
\nonumber\\[1ex]
\!&=&\!2([D^\kappa,D_\mu]+D_\mu D^\kappa)\,\text{tr}(\cF_{\kappa\nu})\,\epsilon^{\alpha\beta\mu\nu} \ .
\eea
Using the combinatorial identity $R_{\mu\alpha\beta\gamma}\epsilon^{\alpha\beta\gamma\nu}=0$, the second piece of (\ref{third piece}) can be dropped after integration by parts
\bea
&&\int d^4 x\ D_\alpha \xi_\beta\, D_\mu D^\kappa \,\text{tr}(\cF_{\kappa\nu})\,\epsilon^{\alpha\beta\mu\nu}
\nonumber\\[1ex]
&&=\,-\int d^4 x\ D_\mu D_\alpha \xi_\beta\, D^\kappa \,\text{tr}(\cF_{\kappa\nu})\,\epsilon^{\alpha\beta\mu\nu} \,=\, 0 \ .
\eea
The other, commutator term can be simplified with the curvature tensor,
\bea
2[D^\kappa,D_\mu]\,\text{tr}(\cF_{\kappa\nu})\,\epsilon^{\alpha\beta\mu\nu}
\!&=&\!2\bigg(\!-R\indices{^\lambda_\kappa^\kappa_\mu}\,\text{tr}(\cF_{\lambda\nu})
-R\indices{^\lambda_\nu^\kappa_\mu}\,\text{tr}(\cF_{\kappa\lambda})\bigg)\,\epsilon^{\alpha\beta\mu\nu}
\nonumber\\[1ex]
\!&=&\!\!\bigg(2R\indices{^\lambda_\mu}\,\text{tr}(\cF_{\lambda\nu})-R\indices{^\kappa^\lambda_\mu_\nu}\,\text{tr}(\cF_{\kappa\lambda})\bigg)\,\epsilon^{\alpha\beta\mu\nu}
\ ,
\eea
bringing us to
\bea
\frac{1}{384\pi^2}\int d^4 x\ D_{\alpha}\xi_{\beta}
\bigg(
R\indices{^\kappa^\lambda_\mu_\nu}\,\text{tr}(\cF_{\kappa\lambda})
+R\,\text{tr}(\cF_{\mu\nu})-4R\indices{^\lambda_\mu}\,\text{tr}(\cF_{\lambda\nu})
\bigg)
\epsilon^{\alpha\beta\mu\nu} \ ,
\eea
for the surviving part of (\ref{Kosmann}) in $d=4$.

Comparing this against the standard result of Alvarez-Gaume and Witten gives the desired relation,
\bea
%\lim_{\beta\rightarrow 0} \text{Tr}(\Gamma{\hat{\boldsymbol\delta}}_\xi)\,=\,
\text{Tr}(\Gamma\hat{\boldsymbol\delta}_\xi)
=\text{Tr}\left[\Gamma\left({\boldsymbol\Delta}_\xi-\frac14\hat\xi_V^{ab}\gamma_{ab}\right)\right]
= \frac12\times \text{Tr}(\Gamma{\boldsymbol\Delta}_\xi)\ ,
\eea
as was claimed, with help from another nontrivial identity in $d=4$,
\bea
\bigg(
R\indices{^\kappa^\lambda_\mu_\nu}\,\text{tr}(\cF_{\kappa\lambda})
+R\,\text{tr}(\cF_{\mu\nu})-4R\indices{^\lambda_\mu}\,\text{tr}(\cF_{\lambda\nu})
\bigg)
\epsilon^{\alpha\beta\mu\nu}\,=\,R\indices{^\alpha^\beta_\mu_\nu}\,\text{tr}(\cF_{\rho\sigma})\,\epsilon^{\mu\nu\rho\sigma}\ ,
\eea
also confirmed by brute-force. We performed the computation in $d=4$, where
the pure diffeomorphism anomaly is absent, so the end result is a mixed anomaly
between the diffeomorphism and Abelian gauge transformations.

\end{document}